\documentclass{aa}

\usepackage[utf8]{inputenc}

\usepackage{amsmath}

\usepackage{txfonts}
\usepackage{chemformula}
\usepackage{multirow}

\usepackage{natbib}
\bibpunct{(}{)}{;}{a}{}{,}
\usepackage{hyperref}
\hypersetup{colorlinks = true, 
urlcolor   = blue, 
linkcolor  = blue, 
citecolor  = blue 
}

\begin{document}

   \title{Origin of hydrogen fluoride emission in the Orion Bar\thanks{\textit{Herschel} is an ESA space observatory with science instruments provided by the European-led Principal Investigator consortia and with important participation from NASA.}}
    
   \subtitle{An excellent tracer for CO-dark H$_{2}$ gas clouds}

   \author{\"U.~Kavak\inst{1,2}
          \and F.~F.~S.~van~der~Tak\inst{2,1} 
          \and A.~G.~G.~M.~Tielens \inst{3}
          \and R.~F.~Shipman\inst{2,1}
          }

   \institute{Kapteyn Astronomical Institute, University of Groningen, P.O. Box 800, 9700 AV Groningen, The Netherlands \\
             \email{kavak@astro.rug.nl}
         \and
             SRON Netherlands Institute for Space Research, Landleven 12, 9747 AD Groningen, The Netherlands
         \and   
              Leiden Observatory, Leiden University, PO Box 9513, NL-2300RA, Leiden, the Netherlands
             }

   \date{Received June 15, 2019; accepted July 16, 2019}

 
  \abstract
   {The hydrogen fluoride (HF) molecule is seen in absorption in the interstellar medium (ISM) along many lines of sight. Surprisingly, it is observed in emission toward the Orion Bar, which is an interface between the ionized region around the Orion Trapezium stars and the Orion molecular cloud.}
   {We aim to understand the origin of HF emission in the Orion Bar by comparing its spatial distribution with other tracers. We examine three mechanisms to explain the HF emission: thermal excitation, radiative dust pumping, and chemical pumping.}
   {We used a \textit{Herschel}/HIFI strip map of the HF \textit{J} = 1$\,\to\,$0 line, covering 0.5$^{\arcmin}$ by 1.5$^{\arcmin}$ that is oriented perpendicular to the Orion Bar. We used the RADEX non-local thermodynamic equilibrium (non-LTE) code to construct the HF column density map. We use the Meudon PDR code to explain the morphology of HF.}
   {The bulk of the HF emission at $10\ \mathrm{km~s^{-1}}$ emerges from the CO-dark molecular gas that separates the ionization front from the molecular gas that is deeper in the Orion Bar. The excitation of HF is caused mainly by collisions with H$_{2}$ at a density of $10^{5}\ \mathrm{cm^{-3}}$ together with a small contribution of electrons in the interclump gas of the Orion Bar. Infrared pumping and chemical pumping are not important.}
   {We conclude that the HF \textit{J} = 1$\,\to\,$0 line traces CO-dark molecular gas. Similarly, bright photodissociation regions associated with massive star formation may be responsible for the HF emission observed toward active galactic nuclei.}


   \authorrunning{Kavak et al.} 
   \maketitle
%
\section{Introduction} \label{intro}
The penetration of UV$-$photons ($h\nu < 13.6\ \mathrm{eV}$), emitted by massive stars, leads to bright regions at the edges of molecular clouds that are called photo-dissociation regions (PDRs)\footnote{We prefer this term over photon-dominated region, because HII regions and AGN nuclei are also dominated by photons; however, we use the term photo-ionization of atoms rather than photodissociation of molecules).} \citep{hollenbach99, wolfire03}. PDRs can also be seen in high-mass star-forming regions, protoplanetary disks, and the nuclei of active galaxies. The penetration of FUV photons regulates the thermal and chemical balance of the gas in a PDR. The gradual decrease of the FUV flux in a PDR results in a layered structure \citep{tielens93} where a chemical phase transition, such as H$^+$ ${\rightarrow}$ H ${\rightarrow}$ H$_{2}$ and C$^{+}$ ${\rightarrow}$ C ${\rightarrow}$ CO, occurs \citep{kaufman99,wolfire03}.

The Orion Bar is a prototypical PDR at a distance of $414\ \mathrm{pc}$ \citep{tauber94, menten07}, located between the Orion molecular cloud and the Orion Nebula, the HII region surrounding the Trapezium stars. Observations at infrared and sub-millimeter wavelengths first indicate a geometry for the bar where the PDR is wrapped around the Orion Nebula and second, changes from a face-on to an edge-on view in the Orion Bar where the molecular emission peaks \citep{hogerheijde95, walmsley00}. The mean temperature of the molecular gas in the bar is $85\ \mathrm{K}$, while the temperature rises to several $100\ \mathrm{K}$ toward the ionization front \citep{ossenkopf13}, where the emission from polycyclic aromatic hydrocarbon (PAH) particles and vibrationally excited H$_{2}$ are observed \citep{walmsley00}.

While the temperature structure of the Orion Bar is reasonably well understood \citep{tielens85, ossenkopf13, nagy17}, the same cannot be said about the density structure. The mean density of the molecular gas is $10^{5}\ \mathrm{cm^{-3}}$, but single-dish observations already indicate the presence of random small-scale density variations, usually called "clumps" \citep{hogerheijde95}, which are also seen toward other PDRs \citep{stutzki88, wang93}. While interferometric observations have confirmed the presence of clumps \citep{youngowl00}, the densities of both the clumps and the interclump medium are somewhat uncertain. The interclump medium probably has a density between a few $10^{4}$ and $2\times 10^{5}\ \mathrm{cm^{-3}}$ \citep{simon97}, while estimates of the clump density range from $1.5\times 10^{6}\ \mathrm{cm^{-3}}$ to $6\times 10^{6}\ \mathrm{cm^{-3}}$ \citep{lisSchilke03}. \citet{javier16} show the presence of even denser and small gas clumps that are close to the edge of the cloud using high-resolution Atacama Large Millimeter Array (ALMA) observations. 

In addition to gas clumps, dust condensations in the Orion Bar were found by \citet{qui18}. These condensations have temperatures between $50-73\ \mathrm{K}$ and masses of between $0.03-0.3\ \mathrm{M_\sun}$, and are very compact, that is, $r < 0.01\ \mathrm{pc}$. They are located right behind the PAH ridge of the Orion Bar. 

We study the origin of the HF emission in the Orion Bar by using a map of the HF \textit{J} = 1$\,\to\,$0 line. We also investigate whether we can use HF as a tracer of CO-dark molecular gas or not. HF is an F-bearing hydride molecule which has been established as a surrogate tracer of molecular hydrogen in diffuse clouds \citep{emprechtinger12}. Halogen-containing molecules like HF have a unique thermochemistry \citep{neufeld09}. In particular, only fluorine has a higher affinity to hydrogen than hydrogen itself so that the reaction,

\begin{center}
\ch{H2 + F -> HF + H,}
\end{center}
is exothermic. Models by \citet{neufeld09} predict that, in the presence of H$_{2}$, all of the gas phase fluorine is rapidly converted into HF, resulting in an abundance of $\sim$2 $\times 10^{-8}$ in diffuse clouds, that is, they are close to the Solar fluorine abundance \citep{neufeld10}. \textit{Herschel} observations of the HF \textit{J} = 0$\,\to\,$1 line confirm this prediction: the line is seen in absorption toward several background sources, with abundances of $\sim$2--3 $\times 10^{-8}$ \citep{neufeld10}. Toward dense clouds, the HF abundance is measured to be $\sim$100 times lower \citep{phillips10}, suggesting significant depletion of F on grain surfaces. In PDRs, the destruction of HF occurs by photo-dissociation \citep{neufeld97} at a rate of $1.17\times 10^{-10}\ \mathrm{s^{-1}}$~$\chi_\mathrm{UV}$, where $\chi_\mathrm{UV}$ is the mean intensity of the radiation field that is normalized with respect to the standard interstellar UV-radiation field of \citet{draine78}. In addition, reactions with C$^{+}$ can be an important destruction channel \citep{neufeld09}.

HF has been detected in extragalactic sources; such as in emission toward Mrk 231 \citep{vanderwerf10}, as a P Cygni profile toward Arp 220 \citep{rangwala11}, and in absorption toward nearby luminous galaxies \citep{monje14} as well as the Cloverleaf quasar at $z = 2.56$ \citep{monje11_z}. The ground state transition of HF, that is, \textit{J} = 0$\,\to\,$1 appears in absorption in many Galactic lines of sight \citep{neufeld97, neufeld10, sonnentrucker10, monje11, emprechtinger12, vanderwiel16}. In contrast, IRC$+$10216, a well-known Galactic asymptotic giant branch star, shows HF in emission \citep{Agundez2011}. The large dipole moment of HF and the high frequency of its ground state transition indicate that radiative decay to the ground state is swift. At the low densities of the diffuse ISM, most of the HF is in the rotational ground state and emission would be very weak. This explains why HF can then be readily detected in absorption toward strong background sources. As an exception, the HF \textit{J} = 1$\,\to\,$0 line is observed in emission in the Orion Bar \citep{floris12a}, which is illuminated by the Trapezium stars. Three hypotheses are suggested to explain the HF emission: thermal excitation by collisions with H$_{2}$ or other species; radiative pumping by warm dust continuum or H$_{2}$ line emission at $\sim$2.5 $\mu$m; or chemical pumping where most HF is formed in excited rotational states. To address this issue, we analyzed a spatial map of the HF emission in the Orion Bar.


We organize the paper as follows. In Section~\ref{observations}, we describe the observations, observing modes, and data reduction. In Section~\ref{results}, we present direct observational results, while Section~\ref{analysis} consists of the analysis of the data and a comparison of tracers. In Section~\ref{discussion}, we discuss the hypotheses and the most efficient excitation mechanism for the HF emission. Finally, in Section~\ref{conclusion}, we summarize our main conclusions. 

\section{Observation and data reduction}
\label{observations}

\begin{figure}[t]
   \centering
   \includegraphics[width=1.0\hsize]{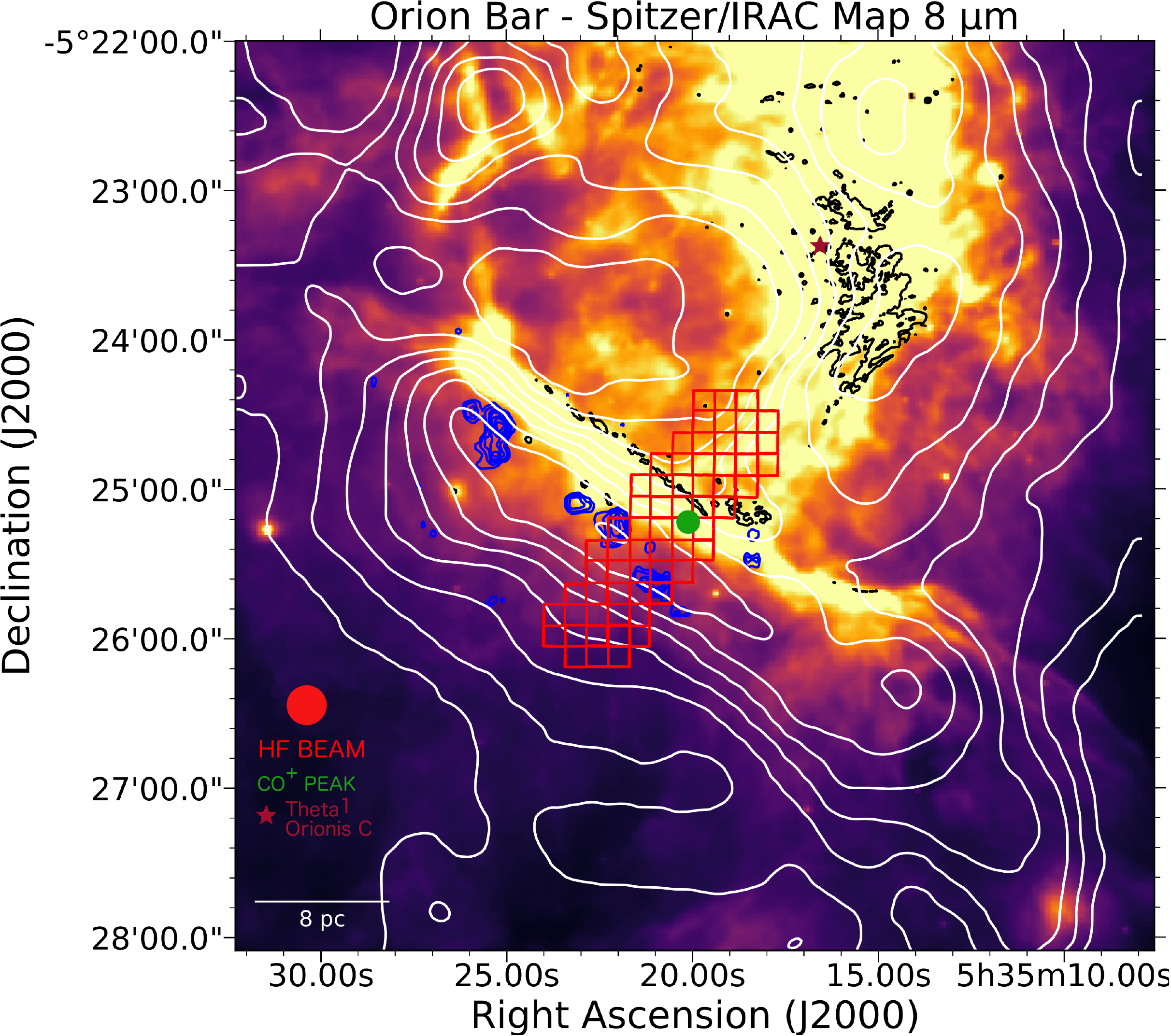}
      \caption{\textit{Spitzer} $8\ \mathrm{\mu}$m map of Orion Bar. Blue contours show H$^{13}$CN \textit{J} = 1${\,\to\,}$0 \citep{lisSchilke03}, which traces dense gas clumps, white contours are $^{12}$CO \textit{J} = 1${\,\to\,}$0 \citep{tauber94}, which traces molecular gas, and black contours are [OI] $6300\ \mathrm{\AA}$ \citep{orionbarOI}, which traces the ionization front. The red squares show the HF strip map perpendicular to the Orion Bar.}
      \label{fig:OrionBar}
   \end{figure}

The observations were made with HIFI (\citealt{degraauw2010}) onboard Herschel \citep{pilbratt10} on 2012 August 28 with observation id (obsid) 1342250409. The area mapped in HF is outlined on emission maps of various molecular tracers in Figure~\ref{fig:OrionBar} assembled on the \textit{Spitzer} 8 $\mu$m map. Receiver 5a was used as the front end for mapping of the Orion Bar in OTF mode, where data are taken continuously while the telescope scans back and forth across the source. In total, one thousand spectra have been obtained. The acousto-optical Wide-Band Spectrometer (WBS) was used as the back-end with full frequency coverage of intermediate frequency (IF) 4~GHz bandwidth in four 1140~MHz sub-bands which have a spectral resolution of 1.1~MHz and a velocity resolution of $1\ \mathrm{km~s^{-1}}$ that is smoothed from the native resolution of $0.2676\ \mathrm{km~s^{-1}}$.

\begin{figure}[ht]
\centering
    \includegraphics[width=0.95\hsize]{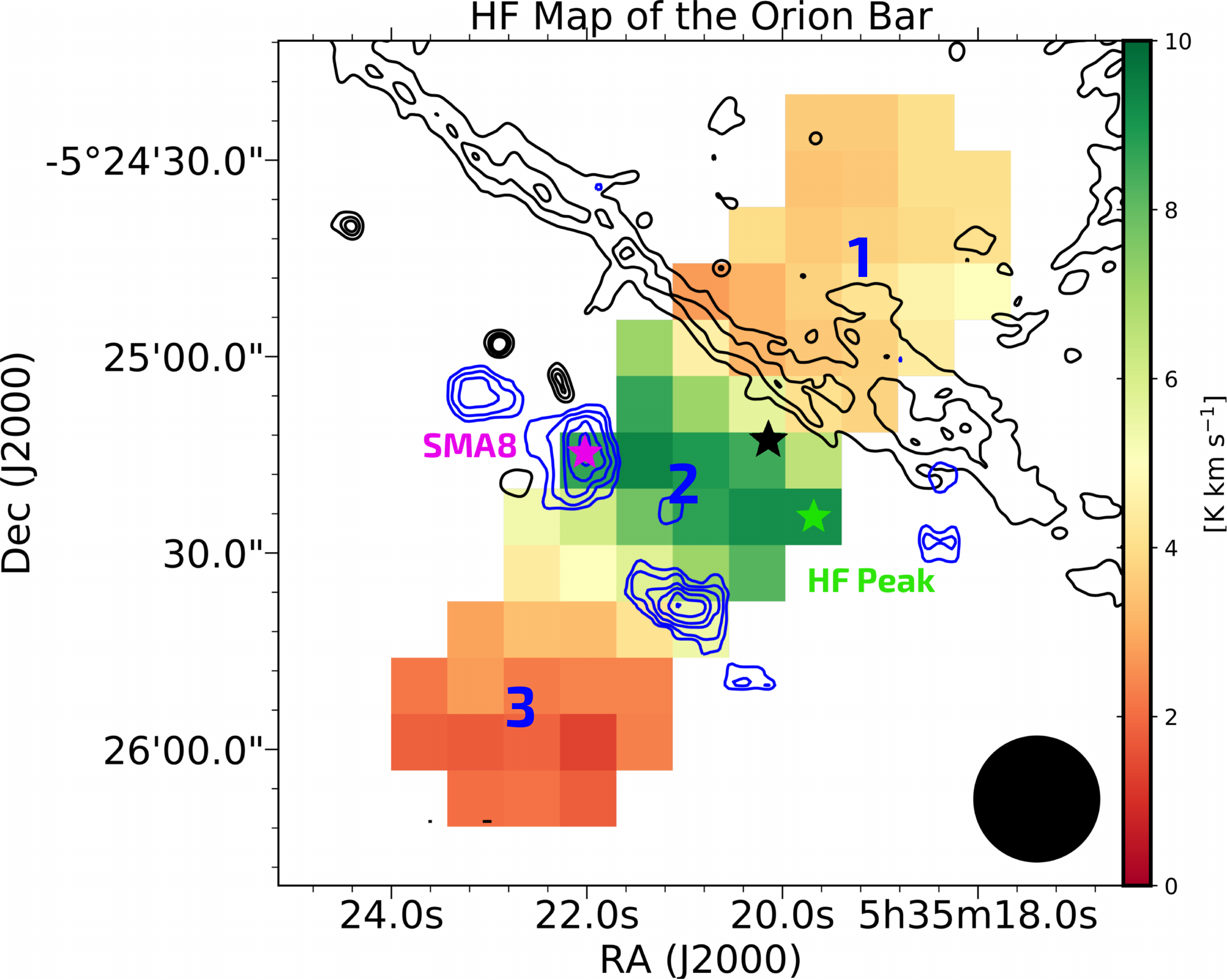} 
\caption{Map of integrated (between $5-13\ \mathrm{km~s^{-1}}$) HF \textit{J} = 1${\,\to\,}$0 intensity overlaid with [OI] $6300\ \mathrm{\AA}$, which traces the ionization front of the Orion Bar and is shown with black contours, and the H$^{13}$CN dense gas tracer, shown in blue contours. The positions where the three spectra in Figure~\ref{fig:3gauss} were extracted are indicated by numbers 1 through 3. The black circle shows the ($18.1^{\arcsec}$) FWHM HIFI beam and the pixel size in this map is $4.5^{\arcsec}$. SMA8 denotes a dust condensation \citep{qui18}. The light green star denotes the HF peak. The black star shows the CO$^+$ peak.}
\label{fig:integratedmap}
\end{figure}

The HF map of the Orion Bar was centered on the CO$^{+}$ peak, that is, $\alpha$ = 05$^\mathrm{h}$35$^\mathrm{m}$20.8$^\mathrm{s}$, $\delta$ = -05$^{\degr}$ 25$^{\arcmin}$ 17.10$^{\arcsec}$ (J2000). Reference spectra have been taken $\sim$5.5$^{\arcmin}$ away at $\alpha$ = 05$^h$35$^m$45.0$^m$, $\delta$ = -05$^{\degr}$ 26$^{\arcmin}$ 16.9$^{\arcsec}$ (J2000). The total integration time (OTF + Reference observation) is 105 minutes. The dobule-sideband system temperature ($T_\mathrm{sys}$) is $920\ \mathrm{K}$. The full width at half maximum (FWHM) beam size at $1232.476\ \mathrm{GHz}$ is $18.1^{\arcsec}$ which corresponds to $7500\ \mathrm{AU}$ or $0.036\ \mathrm{pc}$ at the distance of the Orion Bar.

We inspected the data in the Herschel Interactive Processing Environment \citep[][HIPE]{hipe} version of 15.0.0 for both polarizations. The level 2 data, produced by HIFI-pipeline \citep{russ17}, were exported as a FITS file for further processing in CLASS, which is a sub-package of GILDAS \citep{Gildas2013}. We have estimated the baseline by using a second degree polynomial fit over the entire channel range. After that, we have converted the intensity scale to $T_\mathrm{mb}$ using the mean beam efficiency of 64\% provided by \citet{roelfsema12} to obtain the line parameters. Finally, we have created an integrated intensity map over the $5-13\ \mathrm{km~s^{-1}}$ range. The data cube is the combination of individual spectra at each position.

\section{Results} \label{results}

Fig.~\ref{fig:integratedmap} shows our HF integrated intensity map of the Orion Bar. The HF emission appears as a bright ridge separating the ionization front -- traced by [OI] $6300\ \mathrm{\AA}$ \citep{orionbarOI} -- and the dense molecular clumps -- traced by H$^{13}$CN \textit{J} = 1$\,\to\,$0 \citep{lisSchilke03} -- deeper in the PDR (Fig.~\ref{fig:integratedmap}). Faint HF emission is also observed toward the HII region and the molecular cloud, where we note that the former is brighter than the latter.

\begin{table}[ht]
\caption{Parameters of Gaussian fits in Figure~\ref{fig:3gauss}.} 
\label{table:3gauss}
\centering                  
\begin{tabular}{c c c c c c}       
\hline\hline            
Position   & $V_\mathrm{LSR}$  & $\int$ $T_\mathrm{mb}\Delta$V  & $\Delta$V      & $T_\mathrm{mb}$ \\
No      & (km s$^{-1}$)     &  (K km s$^{-1}$)               &  (km s$^{-1}$) & (K)   \\
\hline
1       & 8.5 (0.1)   &  3.7 (0.1)    & 3.6 (0.1)          & 1.05  \\
2       & 10.2 (0.1)  &  8.5 (0.2)    & 4.4 (0.1)          & 1.85  \\
3       & 10.1 (0.1)  &  2.4 (0.2)    & 3.8 (0.3)          & 0.59  \\
\hline
\end{tabular}
\end{table}

\begin{figure}[t]
   \centering
   \includegraphics[width=0.9\hsize]{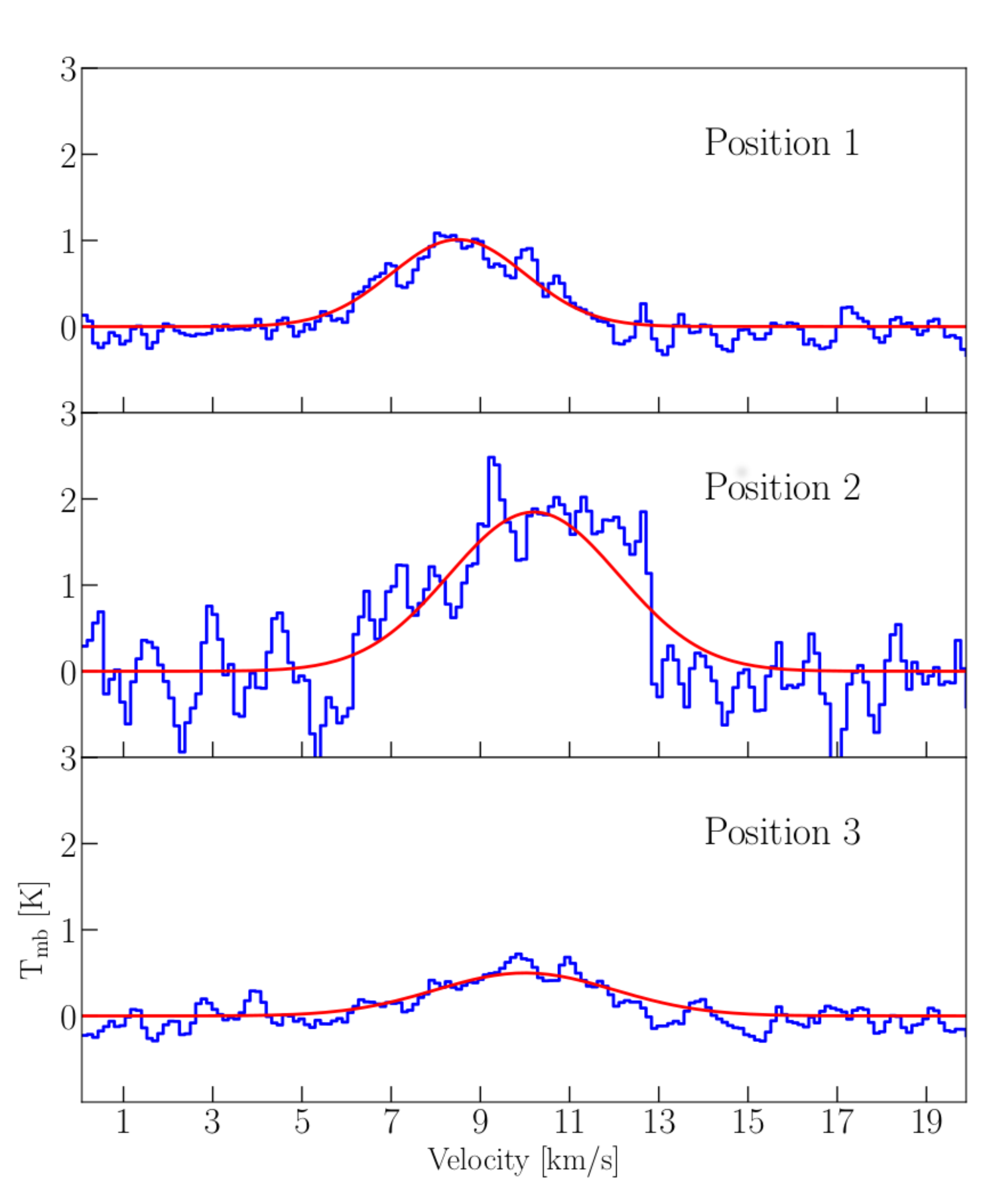}
      \caption{Upper panel shows HF spectrum toward HII region at position 1 and Gaussian fit, which is in red. The middle panel, position 2, shows the spectrum at HF peak, which has also been studied by \citet{floris12a}. The components of HF lines is given in Figure~\ref{fig:OrionSketch}. Finally, the bottom panel, position 3, shows the spectrum observed toward the molecular cloud.}
\label{fig:3gauss}
\end{figure}

We inspected all the lines in the data cube and find 3 distinct regions (position 1, 2, and 3 in Figure~\ref{fig:3gauss}) that are representative of the emission in the regions (see Table~\ref{table:3gauss} for the line parameters). We do not see evidence for the weak absorption feature detected by \citet{floris12a} at $5.5\ \mathrm{km~s^{-1}}$ -- and ascribed by them to absorption by foreground atomic gas -- presumably because of the more limited signal to noise ratio (S/N) in our data, that has been revealed by \citet{vanderwerf2013} in the HI counterpart. The strongest absorption feature peaks at $5\ \mathrm{km~s^{-1}}$ that is a few $\mathrm{km~s^{-1}}$ broad. Position 1, toward the HII region (top panel in Fig.~\ref{fig:3gauss}) reveals an HF emission line peaking at $8.5\ \mathrm{km~s^{-1}}$ and a width of $3.5\ \mathrm{km~s^{-1}}$. The HF profile toward the molecular cloud, position 3, peaks at $10\ \mathrm{km~s^{-1}}$ (Fig.~\ref{table:3gauss}), similar to the main component at the peak of the HF emission, that is, position 2.

\begin{figure}[ht]
\centering
    \includegraphics[width = 0.46\textwidth]{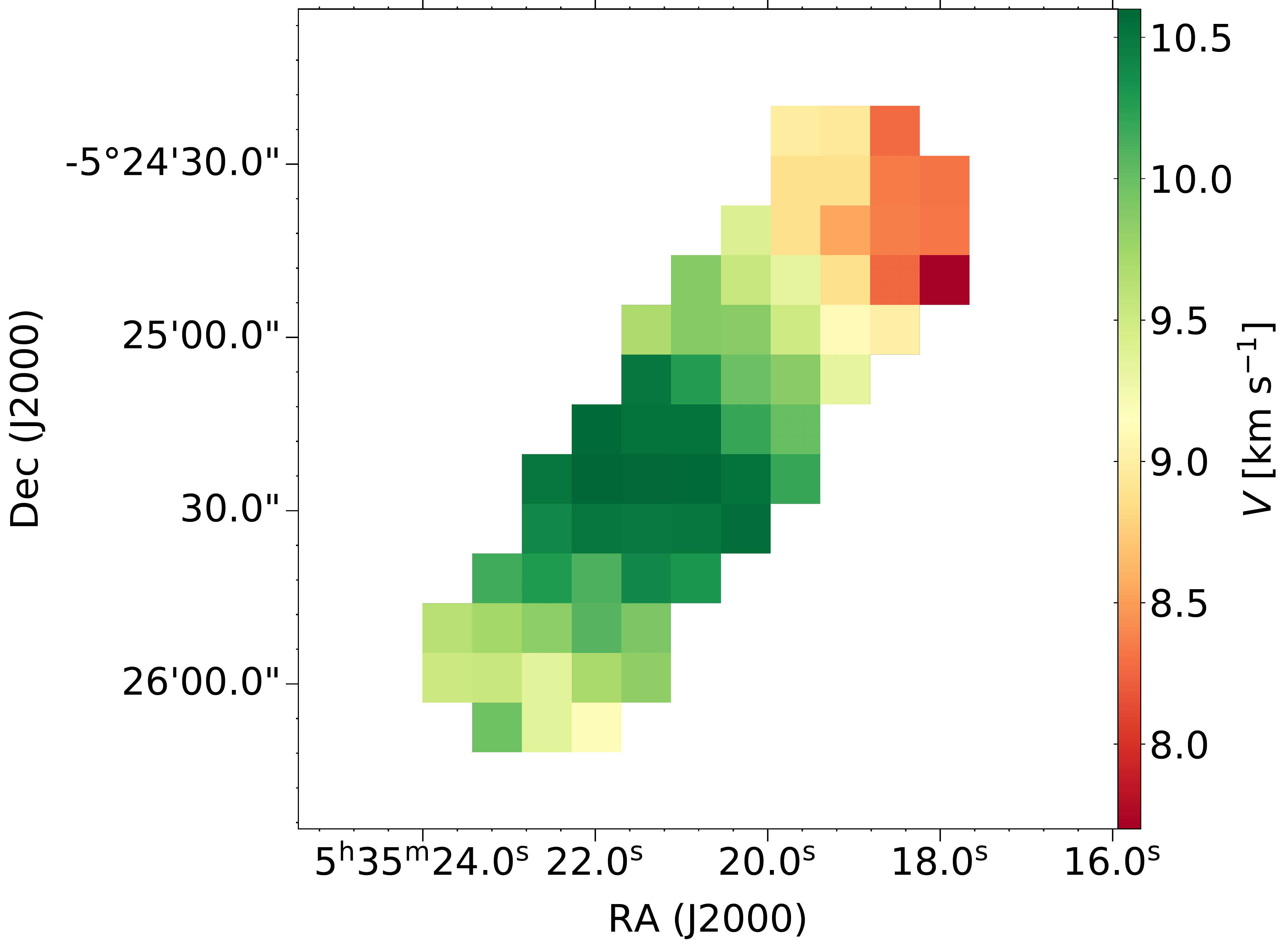}
    \includegraphics[width = 0.45\textwidth]{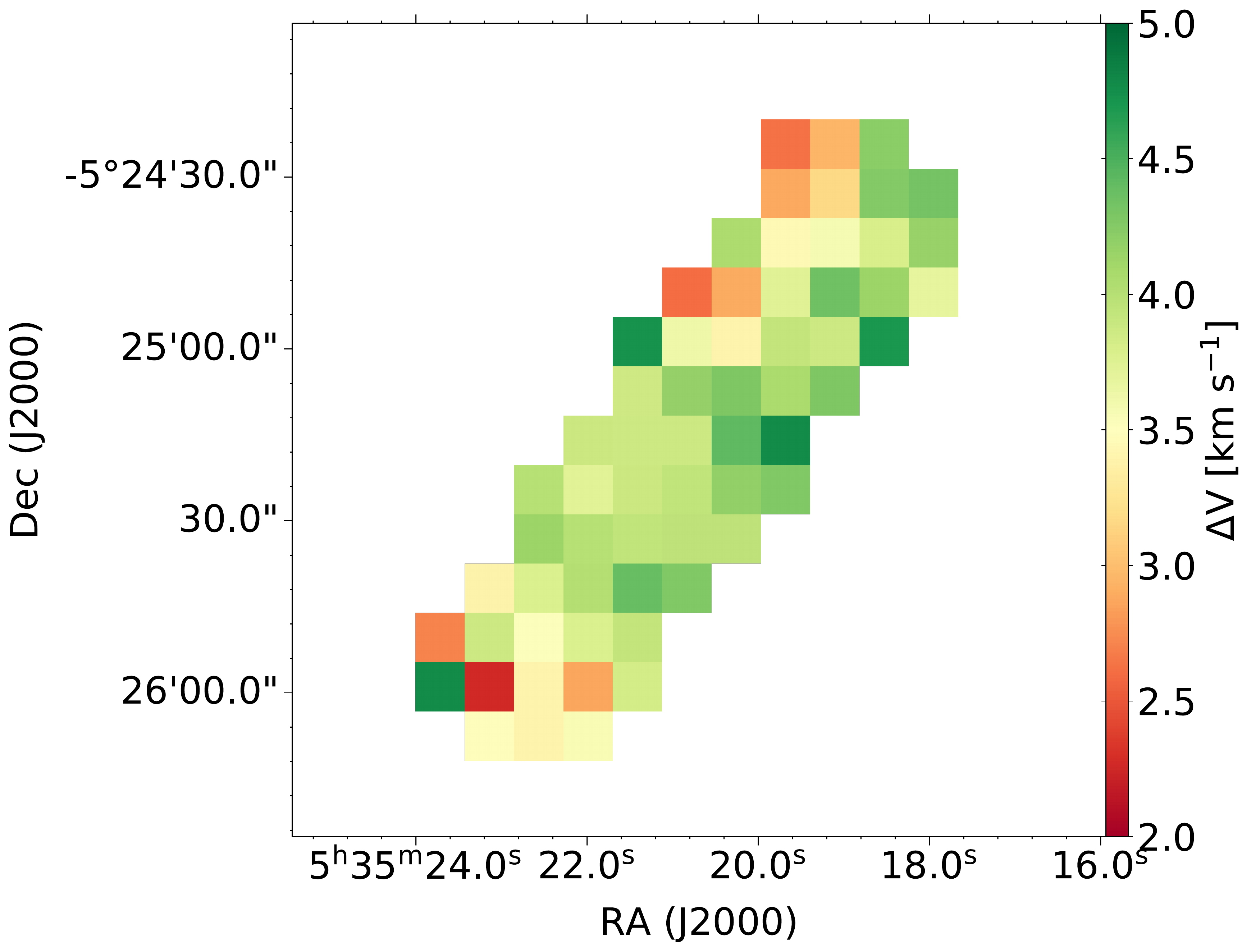}
\caption{\textit{Upper panel}: The map of the central velocity of the HF. There are two velocity component of the HF in the strip map. The \textit{V} = $10.7\ \mathrm{km~s^{-1}}$ component is moving with the Orion Bar itself since it has same velocity distribution. \textit{Bottom panel}: FWHM map of HF \textit{J} = 1$\,\to\,$0 which represents a distribution of the width of $4\ \mathrm{km~s^{-1}}$.}
\label{fig:HF_VelFWHM}
\end{figure}

The velocity at position 1 corresponds to the velocity of the [C\,{\sc ii}] 158 $\mu$m line ($9\ \mathrm{km~s^{-1}}$) rather than the CO background gas ($10\ \mathrm{km~s^{-1}}$; \citet{pabst19}). Hence, the HF emission originates in the PDR evaporative flow from the background molecular cloud as traced by the [C\,{\sc ii}] emission. The typical width of the HF emission is $\sim$4\ $\mathrm{km~s^{-1}}$ and does not vary systematically with position across our map (see Figure~\ref{fig:HF_VelFWHM}). Hence, the HF emission is likely associated with interclump gas, which typically has $\sim$4--5\ $\mathrm{km~s^{-1}}$ wide emission lines \citep{nagy13}. In contrast, the width of emission lines originating in the dense clumps is typically $\sim$2--3\ $\mathrm{km~s^{-1}}$.

\section{Analysis}\label{analysis}
The HF \textit{J} = 1$\,\to\,$0 transition has a critical density ($10^{9}~\mathrm{cm^{-3}}$) much higher than the gas density ($10^{5}~\mathrm{cm^{-3}}$) in the Orion Bar. Thus the HF line is sub-critically excited, and hence the derived column density and abundance are sensitive to physical conditions, that are, density ($n$) and temperature ($T$). Therefore, we have modeled the HF lines to determine the column density.

\subsection{Column density}\label{subsec:columndensity}
We used the RADEX non-LTE radiative transfer code that has been developed to infer physical parameters such as temperature and density, based on statical equilibrium calculations \citep{floris07}. RADEX is available for public use as part of the Leiden Atomic and Molecular Database \citep[LAMDA;][]{schoier05}. The input parameters are kinetic temperature ($T_\mathrm{kin}$), gas density ($n_\mathrm{H_2}$), and molecular column density ($N_\mathrm{col}$). In addition, the FWHM of the line, collisional partners and their collisional data, and radiation field (CMB with or without dust emission) have to be specified as input parameters.

We consider three collision partners for the RADEX models, namely atomic H, H$_{2}$, and electrons. We use the new rate coefficients for the HF-H system by \citet{hf_fdata} which are provided between 10 and $500\ \mathrm{K}$. \citet{yang15} published rate coefficients for p-H$_{2}$ with HF for temperatures up to 3000~K. The previous coefficients for the HF-H$_{2}$ system provided by \citet{guillon12} are consistent with the more recent \citet{yang15} results, and hence we use the coefficients of \citet{guillon12}. Based on quantum mechanical calculations of collisional cross sections for the e-HF system by \citep{thummel92} for $T > 500\ \mathrm{K}$, \citet{floris12a} estimated the excitation rate by electrons for HF $\Delta J$ = 1 at $T < 500\ \mathrm{K}$.

\begin{figure}[h]
\centering
    \includegraphics[width=0.97\hsize]{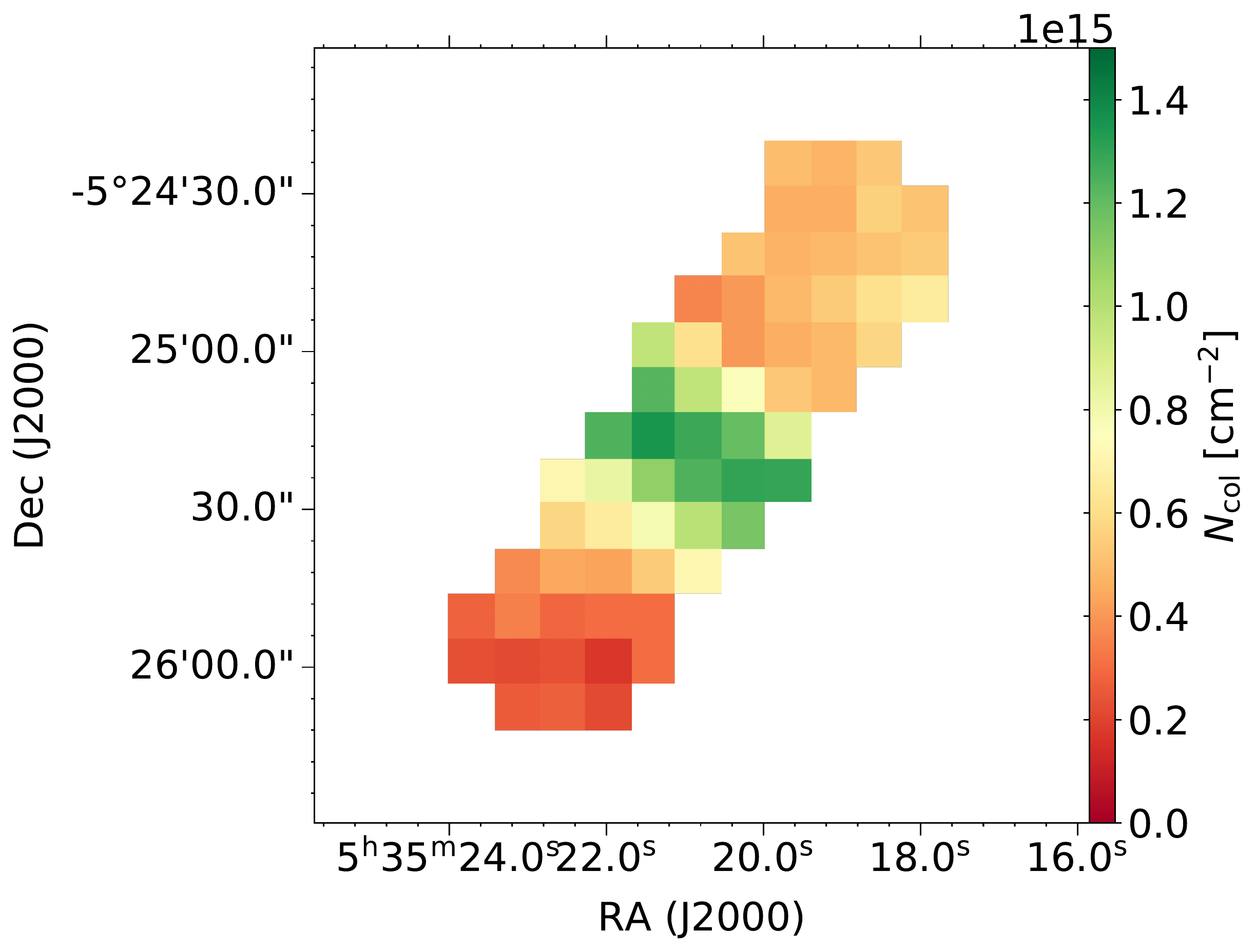}
\caption{The map of the HF column density in the \textit{J} = 1 level. Only cosmic microwave background (CMB) emission is considered as background emission where $T_\mathrm{bg} = 2.73\ \mathrm{K}$.}
\label{fig:Ncolmap_withoutdust}
\end{figure}

For the Orion Bar, we adopt the mean gas temperature as $120\ \mathrm{K}$ \citep{tauber94}, and the density as $10^{5}\ \mathrm{cm^{-3}}$ based on previous observations \citep{floris12a, nagy13}. We calculated the column density at each position in the HF integrated intensity map iteratively to fit the observation for the construction of the column density map in Figure~\ref{fig:Ncolmap_withoutdust} where only CMB emission, $T = 2.73\ \mathrm{K}$, is considered as background emission.

\begin{figure}
\centering
    \includegraphics[width=0.95\hsize]{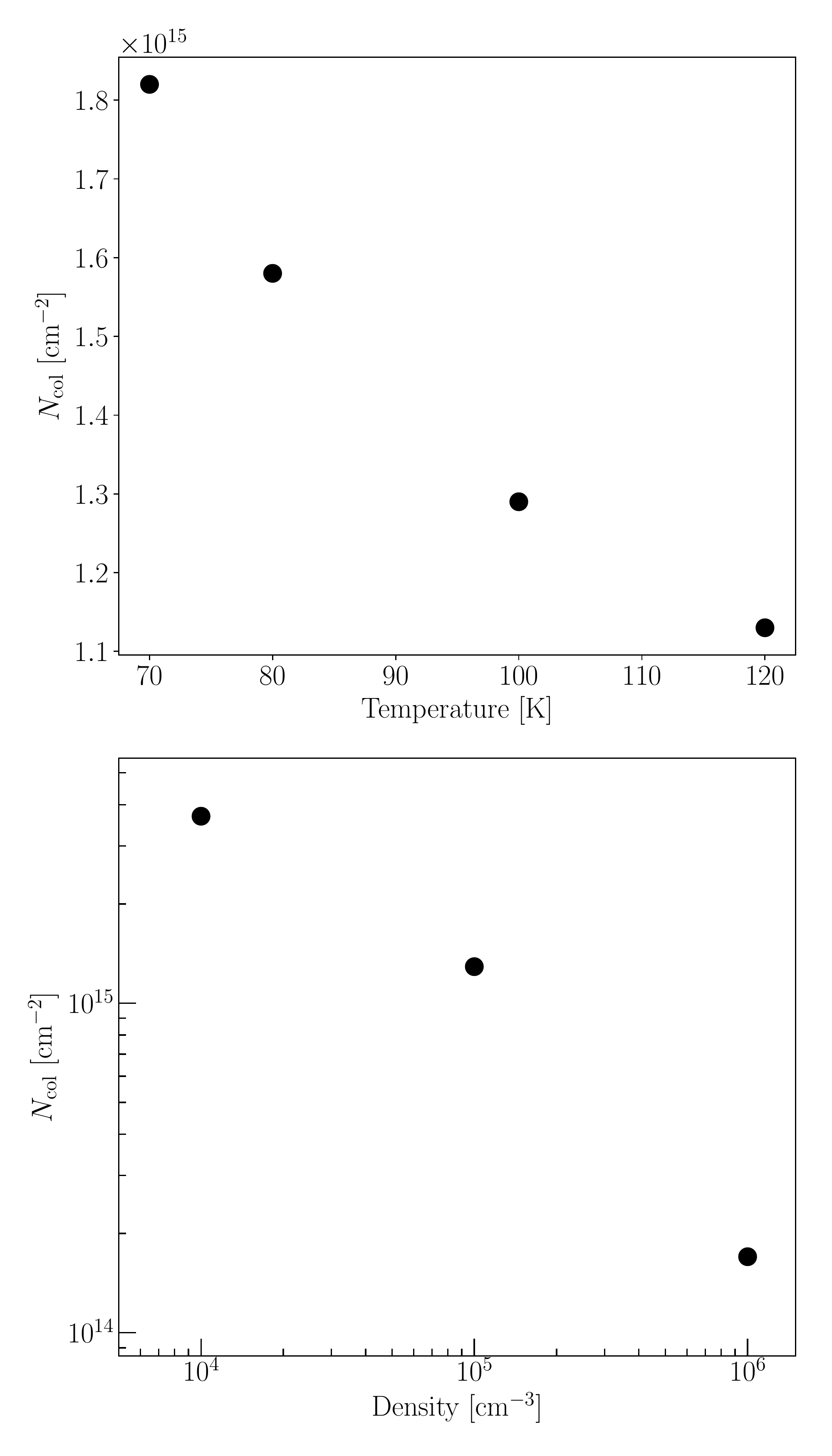}
\caption{Effect of the assumed gas temperature from 70 to $120\ \mathrm{K}$ and H$_{2}$ density from 10$^4$ to 10$^5$~cm$^{-3}$ on the estimated column density of HF based on the RADEX models.}
\label{fig:densityandtemp}
\end{figure}

We have also run models which include a contribution from dust, which has a temperature between $35-70\ \mathrm{K}$ in the Orion Bar \citep{Arab2012}. To that end, we have fitted the observed far-IR dust Spectral Energy Distribution (SED) at different locations (see Appendix~\ref{fig:HFSEDs} for the chosen positions and SEDs) and fitted those with a modified black body (cf., \citet{Arab2012}) and used those parameters to describe the IR radiation field in our RADEX analysis. We have investigated the (excitation) effects of the IR radiation field. To that end we have assembled the IR spectral energy distribution from \textit{Herschel} observations and included this in the RADEX models. The results are insensitive to the IR radiation field because dust is highly optically thin ($\tau \sim 0.02$) at three positions. Hereby, we report in Fig.~\ref{fig:Ncolmap_withoutdust} the results of our models using only the CMB as a background radiation field (see Appendix~\ref{app1} for details). RADEX calculates the optical depth for HF $J$ = 1-0 is 9.6 at N(HF) = 10$^14$ cm$^{-2}$. Our models take line trapping into account as RADEX allow us to quantify this.

Figure~\ref{fig:densityandtemp} shows how variations in the gas temperature and density affect the derived HF column density focusing on the HF peak. The derived column density is inversely proportional to the temperature over the range $70-120\ \mathrm{K}$ (see Figure~\ref{fig:densityandtemp}). However, as the temperature of the gas is much better constrained than the density, the main (systematic) uncertainty in the column density is due to the uncertainty in the density. Given the high critical density of the \textit{J} = 0$\,\to\,$1 line of HF, the derived column density is inversely proportional to the density of the gas over the relevant density range ($10^{4}-5\times10^{6}\ \mathrm{cm^{-3}}$; Figure~\ref{fig:densityandtemp}).

\begin{figure}
\centering
\includegraphics[width=1.0\hsize]{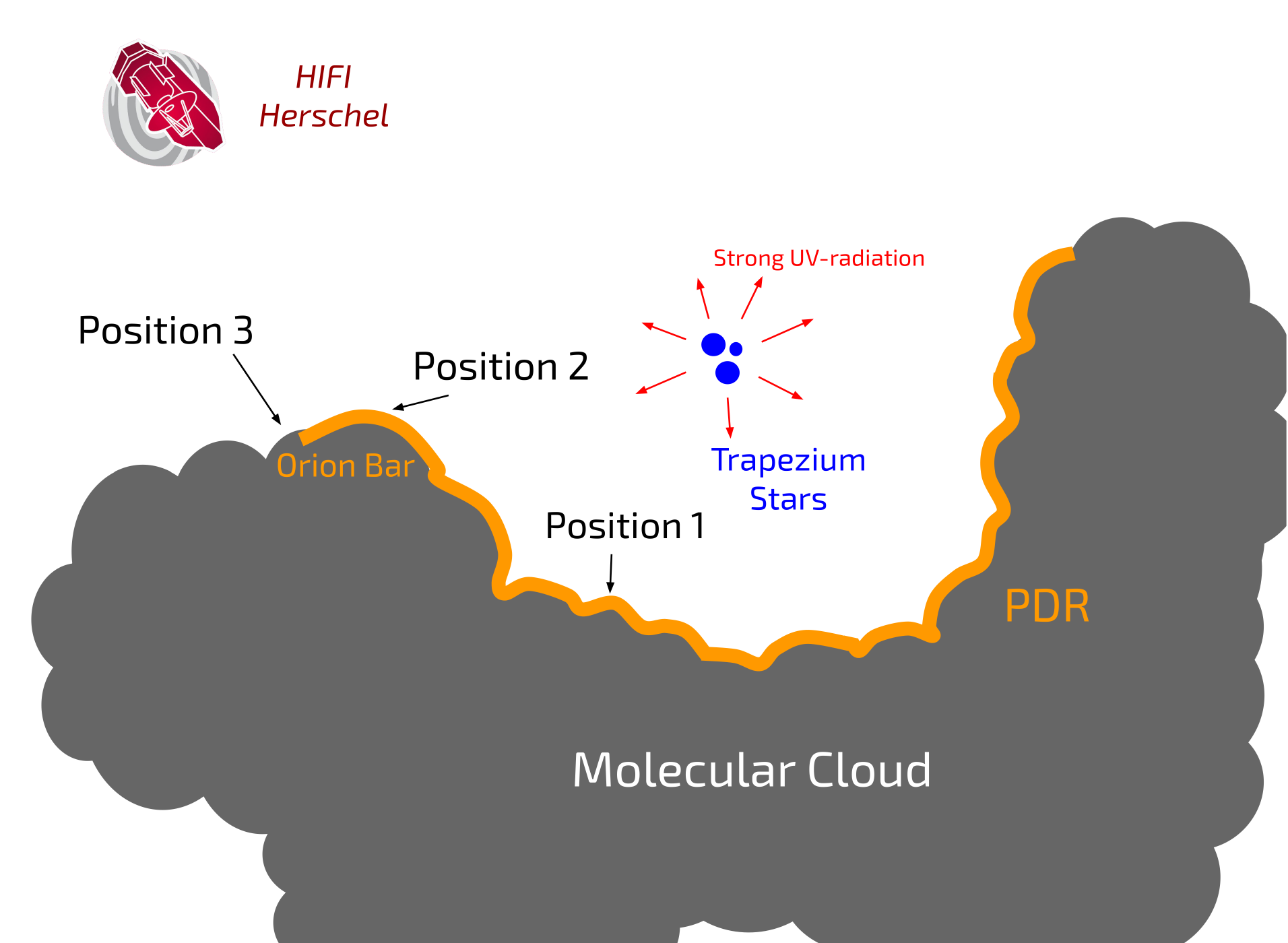}
\caption{Sketch of the Orion Bar. HF emission is observed toward the HII region background molecular cloud originated due to inclination of the Orion Bar. The three example of HF spectra from 3 positions are given in Fig.~\ref{fig:3gauss}. The figure is not to scale.}
\label{fig:OrionSketch}
\end{figure}

\subsection{Spatial distribution of HF}
\begin{figure*}
    \centering
    \includegraphics[width=1.0\hsize]{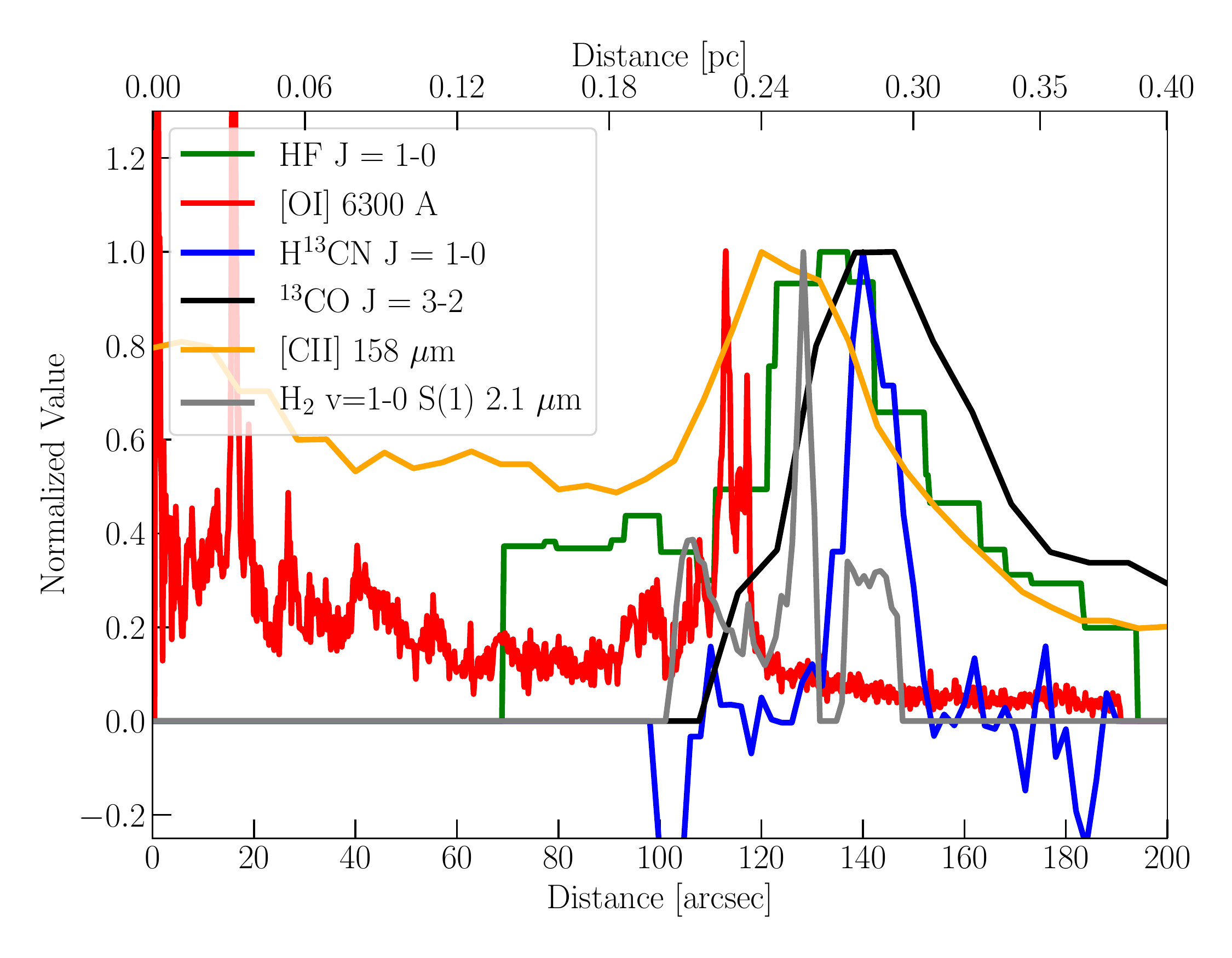}
    \caption{The spatial distribution of different tracers along a crosscut which was chosen over the Orion Bar where the layered structure of the Orion Bar can be seen. The plot starts from $\theta^1$ Ori C which is the main ionizing member of the Trapezium stars. The spatial resolution of HF, [OI], H$^{13}$CN, $^{13}$CO, [C\,{\sc ii}], and H$_2$ is 18.1$^{\arcsec}$, 0.2$^{\arcsec}$, 9.2$^{\arcsec}$, 22$^{\arcsec}$, 11.4$^{\arcsec}$, respectively.}
    \label{fig:tracers}
\end{figure*}

In Figure~\ref{fig:tracers}, we compare the spatial distribution of HF with other species: [O\,{\sc i}] $6300\ \mathrm{\AA}$ \citep{orionbarOI} traces the ionization front, H$^{13}$CN \textit{J} = 1$\,\to\,$0 traces dense clumps in the PDR from \citet{lisSchilke03}, and $^{13}$CO \textit{J} = 3$\,\to\,$2 traces molecular gas in the PDR \citep{tauber94}. For this, we use a crosscut starting from $\theta^1$ Ori C through the HF integrated intensity strip map in Figure~\ref{fig:integratedmap}. We find that the HF emission peaks between the ionization front and the dense molecular gas in the PDR (Fig.~\ref{fig:tracers}). HF has a flat intensity distribution at offsets between 75$^{\arcsec}$ and 100$^{\arcsec}$ toward the HII region while its intensity is decreasing toward the inner part of the molecular cloud. As evidenced by its shifted peak velocity, the emission toward the north west of the strip scan is likely due to the background PDR behind the HII region \citep{salgado16, javier16}. We describe the components of the HF lines with a sketch of the Orion Bar (see Figure~\ref{fig:OrionSketch}). The cross cut in Fig.~\ref{fig:tracers} clearly illustrates that the HF emission straddles the region separating the [C\,{\sc ii}] 158 $\mu$m and the $^{13}$CO \textit{J} = 1$\,\to\,$0 emitting zones.

\section{Discussion} \label{discussion}

In this section, we address the observed morphology of the HF emission in the Orion Bar. For this, we created chemical and excitation models along the strip map.

\subsection{Collisional excitation}

The observed morphology of the HF map reveals a ridge of emission that separates the peak of the H$_{2}$ and the C$^{+}$ emission near the front of the PDR from the molecular emission deeper in. Moreover, the peak of the HF emission is well displaced from the dense clumps traced in H$^{13}$CN. Hence, we attribute the HF emission to the interclump gas with a typical density of $10^{5}\ \mathrm{cm^{-3}}$ and a temperature of $120\ \mathrm{K}$ \citep{tauber94, hogerheijde95}. This is supported by the rather broad ($4\ \mathrm{km~s^{-1}}$) HF line which is characteristic for interclump gas \citep[][see Section~\ref{results}]{nagy13}. To test this hypothesis, we now compare our observations to the results of a PDR model.

\subsubsection{HF abundance}

\begin{figure}[t]
\centering
\includegraphics[width=0.97\hsize]{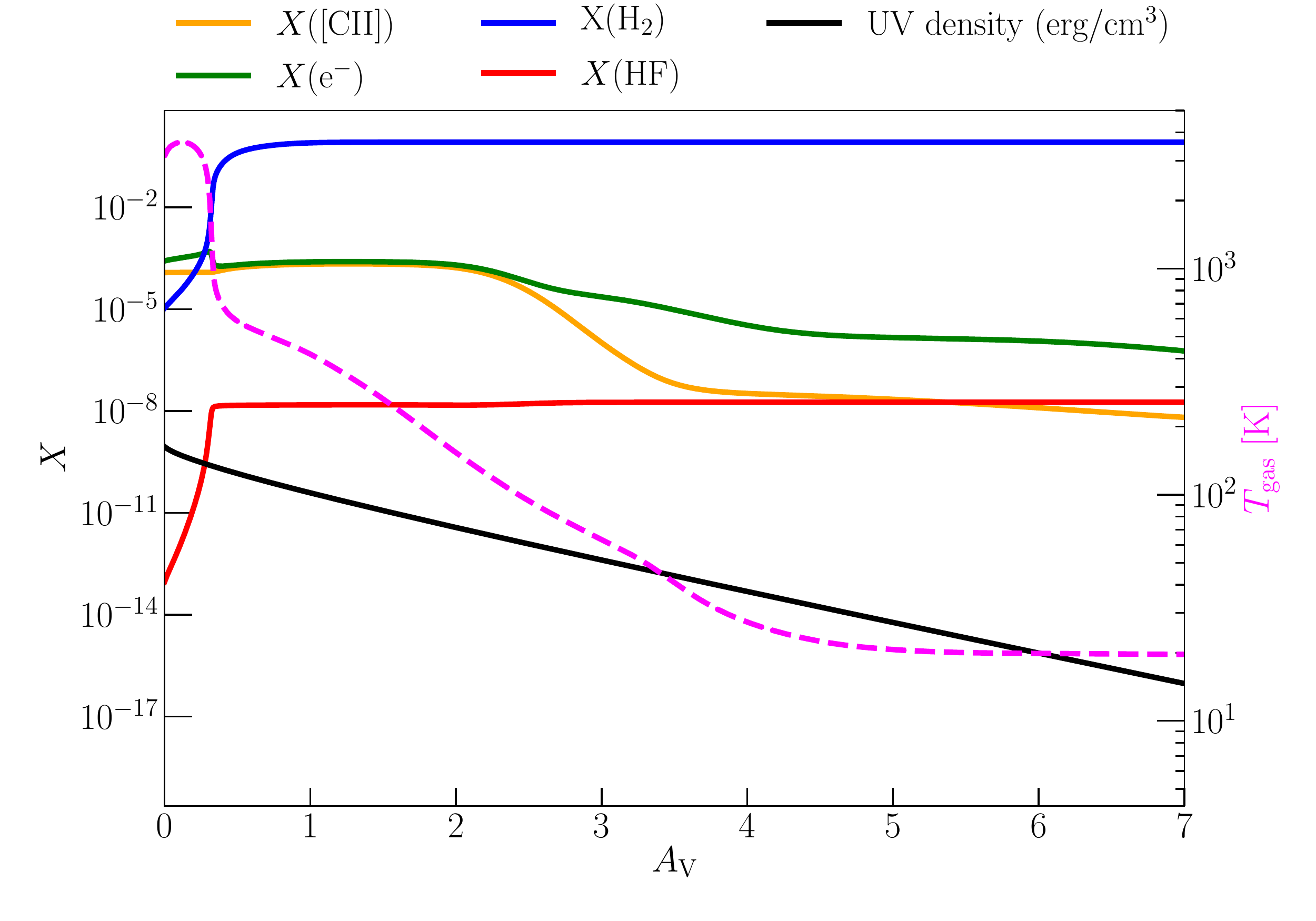} 
\includegraphics[width=0.97\hsize]{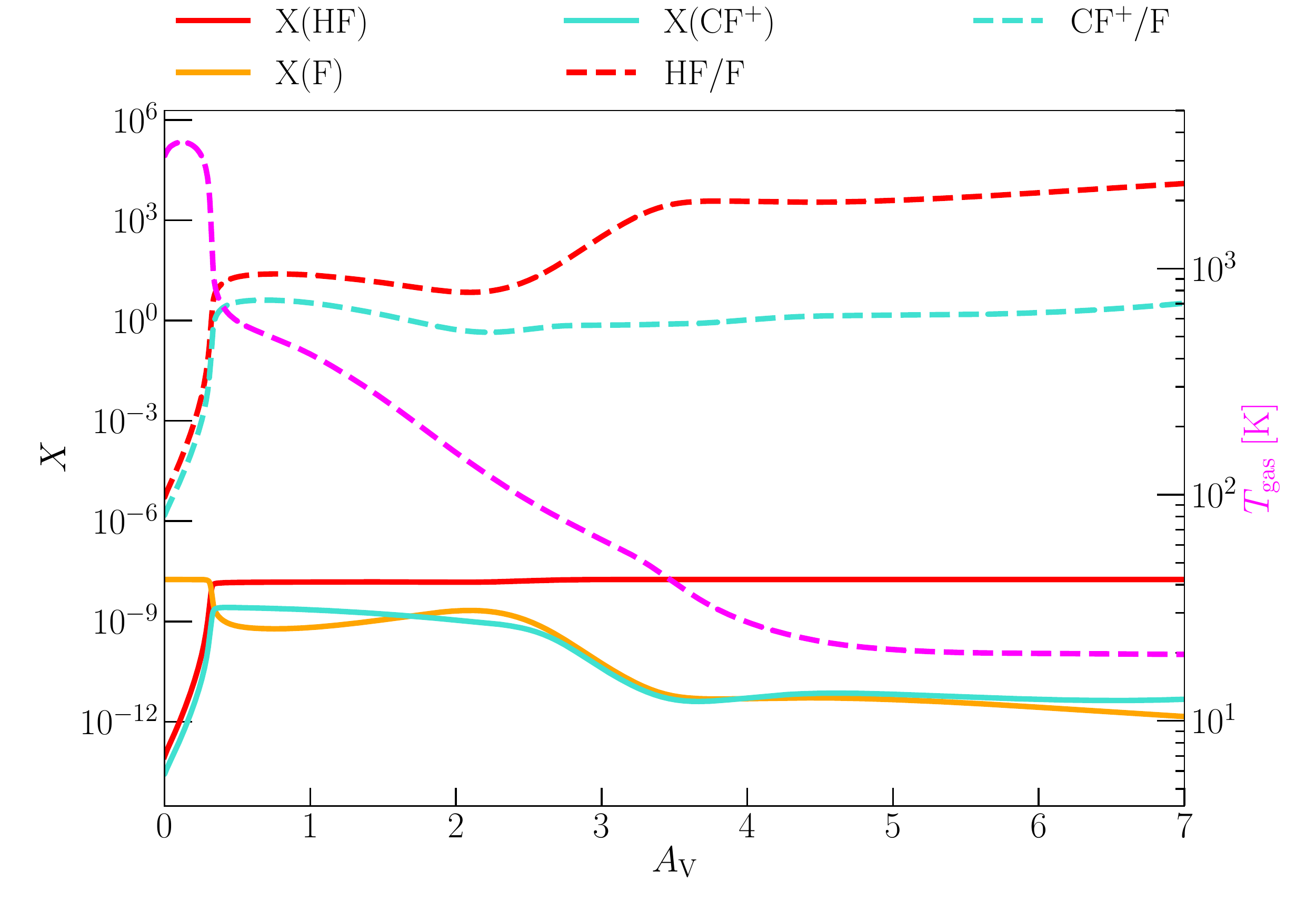}
\caption{\textit{Upper panel:} Abundances of HF, C$^{+}$, H$_{2}$ and electron with UV density corresponding to a Meudon PDR model with a pressure of $P = 10^{8}\ \mathrm{cm^{-3}~K}$. The one illuminated PDR model is considered. The radiation field of $\chi$ = $2.6\times 10^{4}$. \textit{Lower panel:} The abundance of F and the ratio of HF with F and CF$^+$ are given to figure how much of F and CF$^+$ is pushed in to HF. It must be noted that X(F) denotes the abundance of atomic fluorine while in the ratios for the total gas phase fluorine (F $+$ CF$^+$ + HF) abundance. The dashed magenta line shows the gas temperature ($T_\mathrm{gas}$) shown on the right-hand y-axis in both panels.}
\label{fig:abundanceHF-C}
\end{figure}

We have run the Meudon PDR code \citep{muedon06} for a one-dimensional, plane parallel, constant pressure model illuminated on one side by a strong radiation field to determine the spatial distribution of fluorine-bearing species in the PDR. The Meudon code provides the abundances of the major species as a function of depth in the PDR. We have used these results to determine abundances of atomic F, HF, and CF$^+$, using a chemical model \citep{neufeld09}. Specifically, HF is mostly formed in the exothermic reaction of F with H$_{2}$ and destroyed by C$^+$ and UV photons (Fig. \ref{fig:abundanceHF-C}). The dominant reactions playing a role in the HF abundance are:
\begin{center}
\ch{H2 + F -> HF + H} \\
\ch{HF + h$\nu$ -> H + F} \\
\ch{HF + C+ -> CF+ + H} \\
\ch{CF+ + e -> C + F}
\end{center}
The Meudon PDR code calculates self-consistently the temperature for an isobaric model. The results show that the HF abundance increases at the PDR surface between 0 < $A_\mathrm{v}$ < 1 when atomic H is converted into H$_{2}$. HF becomes the major fluorine bearing species at a depth $A_\mathrm{v} >$ 0.5 where it contains $\sim$90\% of the gas phase F; that is, $X$(HF) = $1.8\times 10^{-8}$ relative to H-nuclei (Fig. \ref{fig:abundanceHF-C}).

\begin{figure}[ht]
\centering
\includegraphics[width=1.0\hsize]{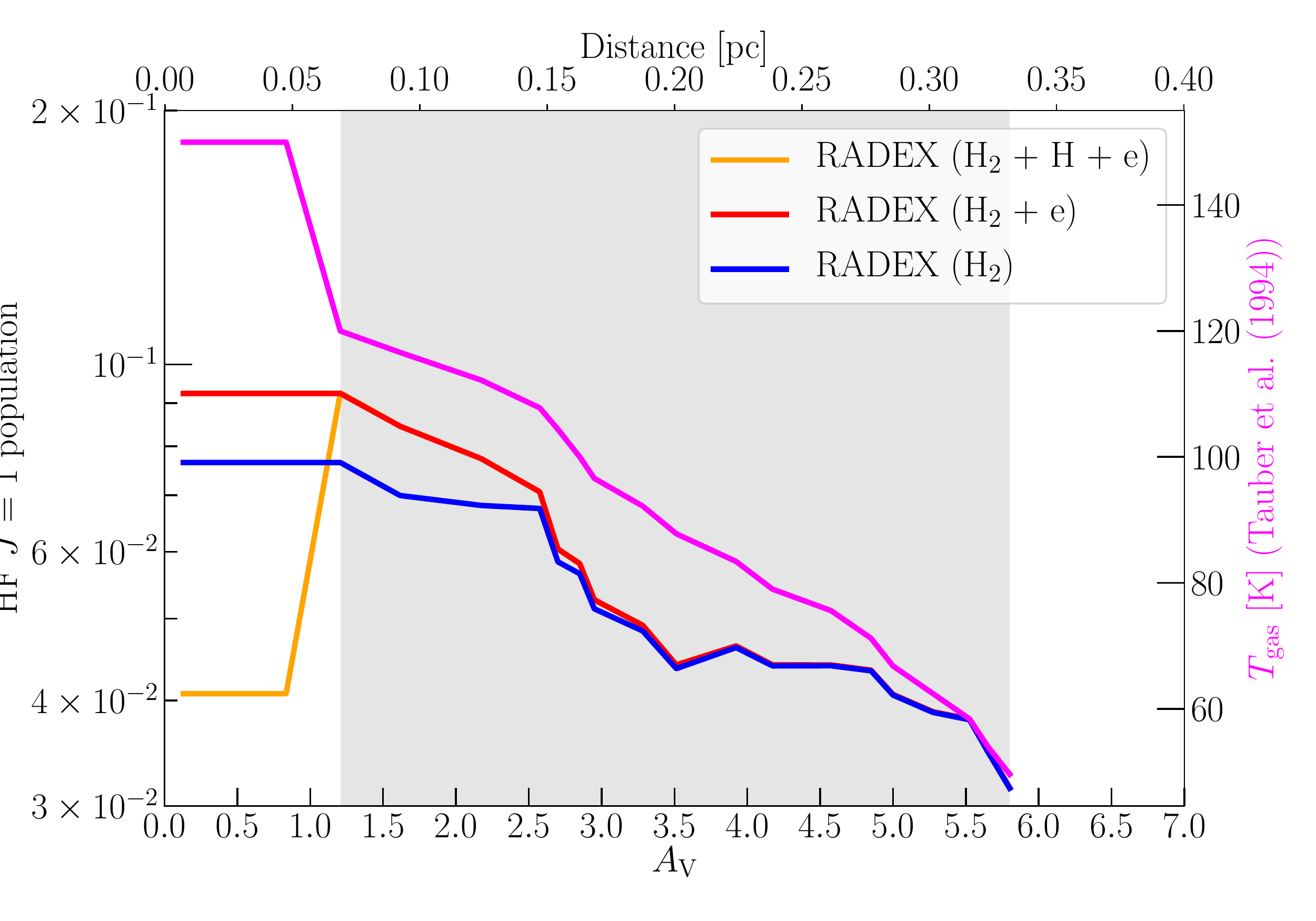}
\caption{HF \textit{J} = 1 level population as a function the depth between $A_\mathrm{v}$= 1.2--6, that is, gray-shaded area. The rest does not reflect proper calculation. The \textit{J} = 1 population is calculated based on the three RADEX models. Blue line shows the model includes only H$_{2}$ as collisional partner. Red curve shows the model consisting of H$_{2}$ and electrons as collisional partners. The model consisting of H$_{2}$, electron, and atomic H does not effect the level population that indicate atomic H is not important for HF excitation at this range. The temperature values shown on right-hand y-axis} are taken from \citet{tauber94}. See the text for the detailed discussion.
\label{fig:J1level_observation}
\end{figure}

Using the calculated H, H$_{2}$, and e abundances from the PDR model, we have calculated the excitation of the \textit{J} $=$ 1 level of HF with RADEX as a function of depth in the PDR (see Fig.~\ref{fig:J1level_observation}). We focus on the range of $A_\mathrm{v}$ of 1.2 and 5.8 as we were only able to extract the gas temperature from $^{12}$CO observation of the Orion Bar \citep{tauber94}. We find that the \textit{J} = 1 level population is typically $0.07$ within this range. This low level population reflects the high critical density of the \textit{J} = 1$\,\to\,$0 transition. The level population is not very sensitive to the H-to-H$_{2}$ conversion near $A_\mathrm{v}$ $=$ 0.5 as both species can readily excite HF \textit{J} $=$ 1. This is a result of a coincidental balancing of the availability of collision partners with their collisional rate coefficients \citep{guillon12, thummel92, hf_fdata, Reese2005}. Deeper in the PDR, the \textit{J} = 1 level population drops. Essentially, this reflects the steep drop in temperature in the model, $T \ll$ $E_{10}$/k as the \textit{J} = 1 level cannot be easily collisionally excited anymore. Anticipating the discussion below, we note that over most of the bright HF emission region of the PDR, excitation is mainly due to collisions with H$_{2}$ with a small ($15\%$) contribution by electrons. Atomic H is not important as a collision partner as H is not abundant in regions where HF is abundant.

Using the PDR model abundance for HF and the excitation results from RADEX, we can calculate the intensity of the HF \textit{J} = 1$\,\to\,$0 line. For this calculation, we have to specify the column density of HF along the line of sight. We adopt a line-of-sight length scale of $0.26\ \mathrm{pc}$, derived by \citet{salgado16} from their analysis of the IR emission from the Orion Bar. With this length scale and our adopted density of H-nuclei, the total column density is $8\times 10^{22}\ \mathrm{cm^{-2}}$. Over much of the PDR, the total column density of HF is thus $8\times 10^{14}\ \mathrm{cm^{-2}}$. The model with N(HF) = $8\times 10^{14}\ \mathrm{cm^{-2}}$ near the peak predicts a line intensity of $1.89\ \mathrm{K}$ at $120\ \mathrm{K}$. We have compared the integrated intensity from RADEX with the observations at the peak of HF. Now, we only need to discuss the drop in intensity deeper in the cloud.

\begin{figure}[ht]
\centering
\includegraphics[width=1.0\hsize]{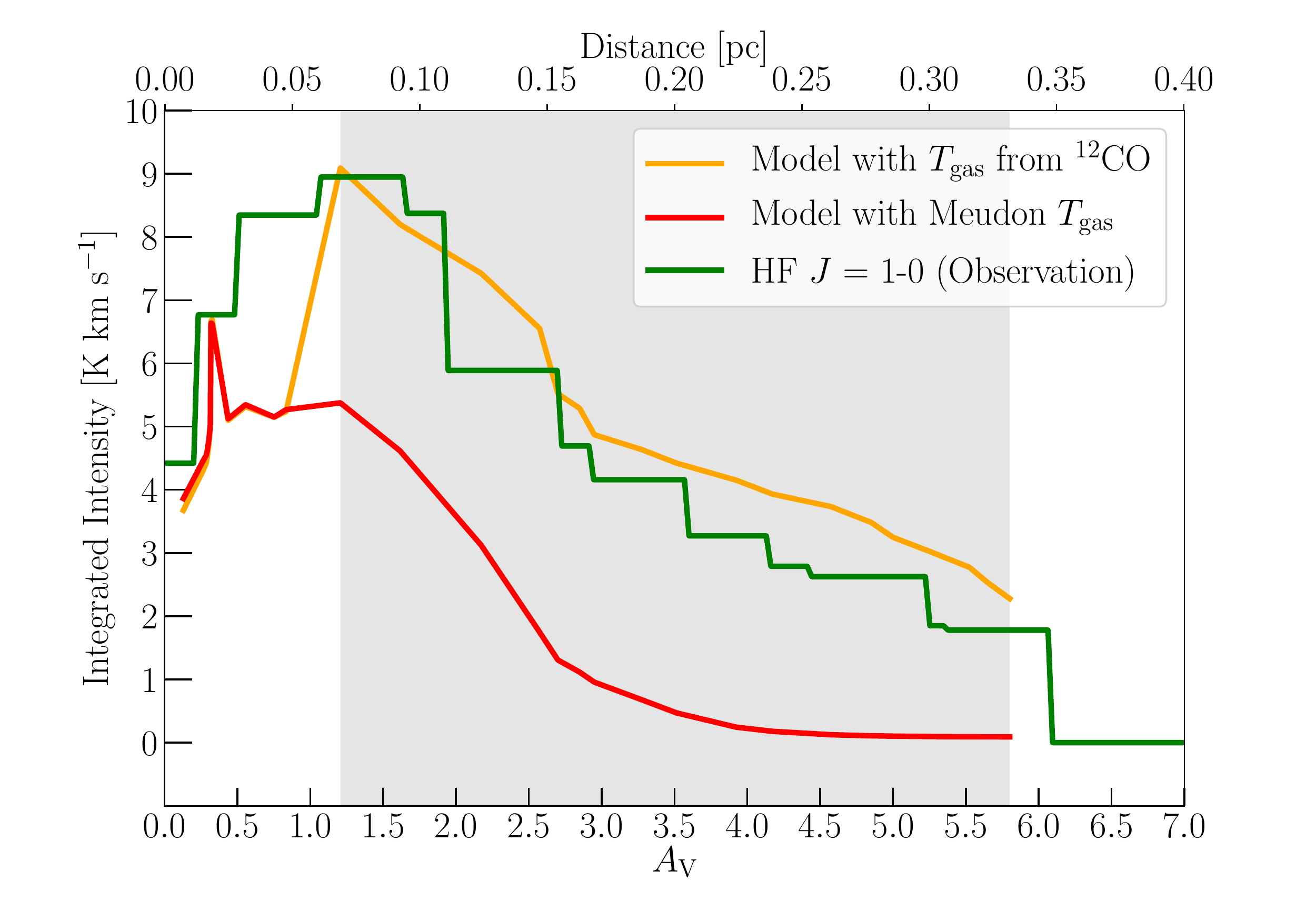}
\caption{Comparison of RADEX models with the HF observation. While green curve shows the HF observation, orange curve show the RADEX model we created with the temperature taken from \citet{tauber94}. Red curve shows a second RADEX model where we use the temperature calculated by Meudon code. We run these models with the same input parameters except for the temperature to figure out the relative importance of the temperature. The temperature is warmer than the model predict in the deep cloud. Since we are unable extract the temperature profile near the surface from $^{12}$CO observations because CO is not formed, we have only focused on the decreasing profile of HF between $A_\mathrm{v}$ = $1.2-5.8$, that is, gray-shaded region, for this comparison. The rest does not reflect proper calculation. See the text for detailed discussion.}
\label{fig:H2_H_e_observation}
\end{figure}

The calculated model intensity distribution is compared to the observations in Figure~\ref{fig:H2_H_e_observation}. With this choice for the HF column density, we reproduce the observed intensity at the peak well. The drop in intensity toward the surface – caused by the drop in HF abundance – is also well reproduced by the model. Fig.~\ref{fig:H2_H_e_observation} shows the comparison of two RADEX models with our observation. However, while the observations show a drop in intensity deep in the cloud, the model underestimates the observed HF intensity. In the model, this drop in intensity is a direct consequence of the steep drop in temperature since the PDR model underestimates the temperature at the surface \citep{Shaw2009, Pellegrini2009}. The calculated temperature, $20\ \mathrm{K}$, is much less than the temperature derived from $^{12}$CO observations, $40\ \mathrm{K}$ \citep{tauber94}. We have calculated a model where we never let the temperature drop below $40\ \mathrm{K}$ (Fig.~\ref{fig:J1level_observation}) and this model reproduces the HF observations well even in the deeper cloud.

Our model reproduces well the observed spatial distribution of the HF emission in the Orion Bar. The ridge of HF emission is an interplay of two factors: the steep rise in the HF abundance when H is converted into HF and the drop in temperature deeper in the PDR when the CO abundance rises and gas cooling is more efficient. Namely, cooling is dominated by CO the deep in the cloud as C$^+$ is not important anymore because C is converted into CO. [OI] cooling is not important as the gas temperature is too low. We conclude therefore that, qualitatively, the HF \textit{J} = 1$\,\to\,$0 line measures the presence of warm dense, CO-dark molecular gas. Quantitatively, the observed intensity is a strong function of the H$_{2}$ density and the column density of HF. We emphasize that the observations measure the HF \textit{J} = 1 column density well. The total HF column density scales then inversely with the adopted density (cf., Fig~\ref{fig:densityandtemp}). Conversely, if we were to fix the total HF column density, then we could adjust the density to reproduce the observed intensity. Our observations cannot break this degeneracy.

\subsection{Infrared pumping}
It has been suggested that the HF line may be excited by infrared photons through the v = 1$\,\to\,$0 fundamental vibrational band at $2.55\ \mu$m given the brightness of the Orion Bar at this wavelength \citep{floris12b}. We compare the vibrational pumping with the collisional excitation of the HF \textit{J} = 1 level. This mechanism is effective if

\begin{equation}\label{infrared_vs_collisional}
    (n_lB_{lu} - n_uB_{ul})J_{near-IR} = n_l n\gamma_{lu}
\end{equation}
where the \textit{B}s are the Einstein coefficients for absorption and stimulated emission, \textit{J$_{ul}$} the mean intensity of the near-IR radiation field, and $\gamma_{lu}$ is the collision probability for pure rotational transitions, which depends on the velocity of molecules in the gas and hence the kinetic temperature. $n_l$ and $n_u$ are the number densities of HF in the lower and upper energy state respectively, and $n$ is the number density of collision partners in the gas. The left-hand side of the equation gives the near-infrared net pumping rate and the right side is the collisional excitation rate. When the left-hand side is greater than the right-hand side, infrared pumping is important. If we ignore stimulated emission as at this low critical density, most of the HF molecule will be in ground state, Eq.~\ref{infrared_vs_collisional} simplifies to, 

\begin{equation}\label{JnearIR}
    J_{near-IR} = (\frac{A_{rot}}{A_{vib}})(\frac{2h\nu^3}{c^2})(\frac{n}{n_{cr}})exp[-h\nu/kT_k].
\end{equation}

We have used the Infrared Space Observatory (\textit{ISO}) Short Wavelength Spectrometer (SWS) spectrum of the Orion Bar \citep{Bertoldi2000}, which is labeled as D8 in the archive\footnote{\href{https://irsa.ipac.caltech.edu/data/SWS/spectra/sws/69501409_sws.tbl}{https://irsa.ipac.caltech.edu/data/SWS/spectra/sws/69501409\_sws.tbl}}. From the spectrum, we estimate the surface brightness of the Bar at $2.55\ \mu$m where the HF vibrational ground state transition lies. The aperture size of SWS is 14${\arcsec}$~$\times$~20${\arcsec}$, and the flux density at the D8 position is $6.16\ \mathrm{Jy}$ which corresponds to a surface brightness of $9.24\times 10^{-14}\ \mathrm{erg~s^{-1}~cm^{-2}~Hz^{-1}~sr^{-1}}$. At $120\ \mathrm{K}$, pumping rate equals $7.38\times 10^{-11}\ \mathrm{s^{-1}}$ from the left side of Eq.~\ref{infrared_vs_collisional}. $\gamma_{01}$ which corresponds to $\gamma_{10}$($g_0$/$g_1$)exp(-$h\nu$/$kT$) that is equal to $4.43\times 10^{-11}\ \mathrm{cm^{3} s^{-1}}$ of the HF molecule where $g_0$ and $g_1$ are the statistical weights of the lower and upper level, respectively. The collisional excitation ($4.43\times 10^{-6}\ \mathrm{s^{-1}}$) is much bigger than the excitation by infrared photons ($7.38\times 10^{-11}\ \mathrm{s^{-1}}$). Therefore, infrared photons do not play a role in the excitation of HF in the Orion Bar.

\subsection{Chemical Pumping}
The third possibility is chemical pumping, where HF is primarily formed in the \textit{J} = 1 or higher states at a reaction rate similar to its radiative decay \citep{floris12b}. To produce HF emission by chemical pumping, the HF formation rate (R = $k_\mathrm{chem}$ $n(\mathrm{H_{2}})~ n(\mathrm{F})$) must equal or exceed the collisional excitation rate of the 1$\,\to\,$0 line. The reaction rate coefficient ($k_\mathrm{chem}$) is equal to $7.78\times 10^{-12}\ \mathrm{cm^{3}~s^{-1}}$ at $120\ \mathrm{K}$ based on \citet{neufeld09}. The density of F is constrained by the total amount of fluorine, $1.8\times 10^{-8}$ relative to H \citep{simondiaz11}, that is, $n(\mathrm{F})$ = $1.8\times 10^{-8} \times n(\mathrm{H_{2}})$ = $1.8\times 10^{-3}\ \mathrm{cm^{-3}}$ where we assumed $n(\mathrm{H_{2}}$) is equal to $1\times 10^{5}\ \mathrm{cm^{-3}}$ in the Orion Bar. Comparison of the chemical pumping rate ($7.78\times 10^{-7}\ \mathrm{s^{-1}}$) with the collisional rate (n$\gamma_{01}$ = $4.43\times 10^{-6}\ \mathrm{s^{-1}}$) for HF \textit{J} = 1$\,\to\,$0 demonstrates that collisional excitation is more important. Chemical pumping does not play a role in the excitation of the HF \textit{J} = 1 level.

\section{Summary} \label{conclusion}

We have determined the most efficient excitation mechanism for HF emission and compared its spatial distribution with other tracers in the Orion Bar. We find that:

\begin{enumerate}
    \item HF emission peaks between the ionization region and the dense gas in the Orion Bar. The line width of HF indicates that HF emission emerges from the interclump medium which has a density of $1\times 10^{5}\ \mathrm{cm^{-3}}$.  

    \item Our model studies shows that the observed peak intensity and the morphology of the emission is well reproduced by collisional excitation by H$_{2}$ molecules with a minor contribution by electrons ($\sim$15\%) while IR pumping or chemical pumping plays no role in its  excitation.

    \item The observations reveal a bright ridge of emission that straddles the boundary between the [C\,{\sc ii}] 158 $\mu$m and the CO emission. This morphology reflects the steep rise of the HF abundance near the surface and the drop in temperature deeper into the PDR. 

    \item The HF \textit{J} = 1 level population peaks in the region where the CO molecule, the common tracer of H$_{2}$, has a low abundance. Such regions are called CO-dark H$_{2}$ gas \citep{madden97, grenier05}. We conclude that HF emission traces CO-dark molecular gas, especially from PDR surfaces, as H$_{2}$ has to be abundant for the formation of HF. In other words, HF \textit{J} = 1$\,\to\,$0 can be used to trace CO-dark H$_{2}$ gas between $A_\mathrm{v}$ = 1.0--3.5 in the Orion Bar. Studies of a wider sample of PDRs will help develop HF as a tracer of CO-dark molecular gas and assist in the interpretation of HF observations of luminous nearby galaxies and high redshift galaxies.
\end{enumerate}

\begin{acknowledgements}
\"U. Kavak wants to dedicate this paper to the memory of Kadir Kangel, one of the biggest supporters of his academic career, who passed away suddenly on 11 May 2019 at the age of 49. We want to thank William Pearson for checking the language of the present paper and Meudon PDR team, especially to Frank Le Petit and Jacques Le Bourlot, for their help with the Meudon code. We also thank Benhui Yang and Benjamin Desrousseaux for sharing their recent collisional data for the HF-H$_{2}$ and HF-H systems. This paper uses \textit{Herschel}-HIFI archival data. HIFI was designed and built by a consortium of institutes and university departments from across Europe, Canada, and the US under the leadership of SRON Netherlands Institute for Space Research, Groningen, The Netherlands, with significant contributions from Germany, France, and the US. Consortium members are Canada: CSA, U.Waterloo; France: IRAP, LAB, LERMA, IRAM; Germany: KOSMA, MPIfR, MPS; Ireland: NUI Maynooth; Italy: ASI, IFSI-INAF, Arcetri-INAF; The Netherlands: SRON, TUD; Poland: CAMK, CBK; Spain: Observatorio Astronomico Nacional (IGN), Centro de Astrobiolog\'{i}a (CSIC-INTA); Sweden: Chalmers University of Technology - MC2, RSS \& GARD, Onsala Space Observatory, Swedish National Space Board, Stockholm University – Stockholm Observatory; Switzerland: ETH Zürich, FHNW; USA: Caltech, JPL, NHSC. HIPE is a joint development by the Herschel Science Ground Segment Consortium, consisting of ESA, the NASA Herschel Science Center, and the HIFI, PACS, and SPIRE consortia. PACS was developed by a consortium of institutes led by MPE (Germany) and including UVIE (Austria); KU Leuven, CSL, IMEC (Belgium); CEA, LAM (France); MPIA (Germany); INAFIFSI/OAA/OAP/OAT, LENS, SISSA (Italy); IAC (Spain).
      
\end{acknowledgements}

\bibliographystyle{aa}
\bibliography{HF_References.bib}

\begin{thebibliography}{60}
\expandafter\ifx\csname natexlab\endcsname\relax\def\natexlab#1{#1}\fi

\bibitem[{Ag{\'{u}}ndez {et~al.}(2011)Ag{\'{u}}ndez, Cernicharo, Waters, Decin,
  Encrenaz, Neufeld, Teyssier, \& Daniel}]{Agundez2011}
Ag{\'{u}}ndez, M., Cernicharo, J., Waters, L. B. F.~M., {et~al.} 2011, A{\&}A,
  533, 6

\bibitem[{{Arab} {et~al.}(2012){Arab}, {Abergel}, {Habart}, {Bernard-Salas},
  {Ayasso}, {Dassas}, {Martin}, \& {White}}]{Arab2012}
{Arab}, H., {Abergel}, A., {Habart}, E., {et~al.} 2012, \aap, 541, A19

\bibitem[{{Bertoldi} {et~al.}(2000){Bertoldi}, {Draine}, {Rosenthal},
  {Timmermann}, {Howat}, {Geballe}, {Feuchtgruber}, \&
  {Drapatz}}]{Bertoldi2000}
{Bertoldi}, F., {Draine}, B.~T., {Rosenthal}, D., {et~al.} 2000, in IAU
  Symposium, Vol. 197, From Molecular Clouds to Planetary, ed. Y.~C. {Minh} \&
  E.~F. {van Dishoeck}, 191

\bibitem[{{de Graauw} {et~al.}(2010){de Graauw}, {Helmich}, {Phillips},
  {Stutzki}, {Caux}, {Whyborn}, {Dieleman}, {Roelfsema}, {Aarts}, {Assendorp},
  {Bachiller}, {Baechtold}, {Barcia}, {Beintema}, {Belitsky}, {Benz}, {Bieber},
  {Boogert}, {Borys}, {Bumble}, {Ca{\"\i}s}, {Caris}, {Cerulli-Irelli},
  {Chattopadhyay}, {Cherednichenko}, {Ciechanowicz}, {Coeur-Joly}, {Comito},
  {Cros}, {de Jonge}, {de Lange}, {Delforges}, {Delorme}, {den Boggende},
  {Desbat}, {Diez-Gonz{\'a}lez}, {di Giorgio}, {Dubbeldam}, {Edwards},
  {Eggens}, {Erickson}, {Evers}, {Fich}, {Finn}, {Franke}, {Gaier}, {Gal},
  {Gao}, {Gallego}, {Gauffre}, {Gill}, {Glenz}, {Golstein}, {Goulooze},
  {Gunsing}, {G{\"u}sten}, {Hartogh}, {Hatch}, {Higgins}, {Honingh}, {Huisman},
  {Jackson}, {Jacobs}, {Jacobs}, {Jarchow}, {Javadi}, {Jellema}, {Justen},
  {Karpov}, {Kasemann}, {Kawamura}, {Keizer}, {Kester}, {Klapwijk}, {Klein},
  {Kollberg}, {Kooi}, {Kooiman}, {Kopf}, {Krause}, {Krieg}, {Kramer},
  {Kruizenga}, {Kuhn}, {Laauwen}, {Lai}, {Larsson}, {Leduc}, {Leinz}, {Lin},
  {Liseau}, {Liu}, {Loose}, {L{\'o}pez-Fernandez}, {Lord}, {Luinge}, {Marston},
  {Mart{\'\i}n-Pintado}, {Maestrini}, {Maiwald}, {McCoey}, {Mehdi}, {Megej},
  {Melchior}, {Meinsma}, {Merkel}, {Michalska}, {Monstein}, {Moratschke},
  {Morris}, {Muller}, {Murphy}, {Naber}, {Natale}, {Nowosielski}, {Nuzzolo},
  {Olberg}, {Olbrich}, {Orfei}, {Orleanski}, {Ossenkopf}, {Peacock}, {Pearson},
  {Peron}, {Phillip-May}, {Piazzo}, {Planesas}, {Rataj}, {Ravera}, {Risacher},
  {Salez}, {Samoska}, {Saraceno}, {Schieder}, {Schlecht}, {Schl{\"o}der},
  {Schm{\"u}lling}, {Schultz}, {Schuster}, {Siebertz}, {Smit}, {Szczerba},
  {Shipman}, {Steinmetz}, {Stern}, {Stokroos}, {Teipen}, {Teyssier}, {Tils},
  {Trappe}, {van Baaren}, {van Leeuwen}, {van de Stadt}, {Visser}, {Wildeman},
  {Wafelbakker}, {Ward}, {Wesselius}, {Wild}, {Wulff}, {Wunsch}, {Tielens},
  {Zaal}, {Zirath}, {Zmuidzinas}, \& {Zwart}}]{degraauw2010}
{de Graauw}, T., {Helmich}, F.~P., {Phillips}, T.~G., {et~al.} 2010, \aap, 518,
  L6

\bibitem[{{Desrousseaux} \& {Lique}(2018)}]{hf_fdata}
{Desrousseaux}, B. \& {Lique}, F. 2018, \mnras, 476, 4719

\bibitem[{{Draine}(1978)}]{draine78}
{Draine}, B.~T. 1978, \apjs, 36, 595

\bibitem[{{Emprechtinger} {et~al.}(2012){Emprechtinger}, {Monje}, {van der
  Tak}, {van der Wiel}, {Lis}, {Neufeld}, {Phillips}, \&
  {Ceccarelli}}]{emprechtinger12}
{Emprechtinger}, M., {Monje}, R.~R., {van der Tak}, F.~F.~S., {et~al.} 2012,
  \apj, 756, 136

\bibitem[{{Gildas Team}(2013)}]{Gildas2013}
{Gildas Team}. 2013, {GILDAS: Grenoble Image and Line Data Analysis Software}

\bibitem[{{Goicoechea} {et~al.}(2016){Goicoechea}, {Pety}, {Cuadrado},
  {Cernicharo}, {Chapillon}, {Fuente}, {Gerin}, {Joblin}, {Marcelino}, \&
  {Pilleri}}]{javier16}
{Goicoechea}, J.~R., {Pety}, J., {Cuadrado}, S., {et~al.} 2016, \nat, 537, 207

\bibitem[{{Grenier} {et~al.}(2005){Grenier}, {Casandjian}, \&
  {Terrier}}]{grenier05}
{Grenier}, I.~A., {Casandjian}, J.-M., \& {Terrier}, R. 2005, Science, 307,
  1292

\bibitem[{{Guillon} \& {Stoecklin}(2012)}]{guillon12}
{Guillon}, G. \& {Stoecklin}, T. 2012, \mnras, 420, 579

\bibitem[{{Herschel Science Ground Segment Consortium}(2011)}]{hipe}
{Herschel Science Ground Segment Consortium}. 2011, {HIPE: Herschel Interactive
  Processing Environment}, Astrophysics Source Code Library

\bibitem[{{Hogerheijde} {et~al.}(1995){Hogerheijde}, {Jansen}, \& {van
  Dishoeck}}]{hogerheijde95}
{Hogerheijde}, M.~R., {Jansen}, D.~J., \& {van Dishoeck}, E.~F. 1995, \aap,
  294, 792

\bibitem[{{Hollenbach} \& {Tielens}(1999)}]{hollenbach99}
{Hollenbach}, D.~J. \& {Tielens}, A.~G.~G.~M. 1999, Reviews of Modern Physics,
  71, 173

\bibitem[{{Kaufman} {et~al.}(1999){Kaufman}, {Wolfire}, {Hollenbach}, \&
  {Luhman}}]{kaufman99}
{Kaufman}, M.~J., {Wolfire}, M.~G., {Hollenbach}, D.~J., \& {Luhman}, M.~L.
  1999, \apj, 527, 795

\bibitem[{{Le Petit} {et~al.}(2006){Le Petit}, {Nehm{\'e}}, {Le Bourlot}, \&
  {Roueff}}]{muedon06}
{Le Petit}, F., {Nehm{\'e}}, C., {Le Bourlot}, J., \& {Roueff}, E. 2006, \apjs,
  164, 506

\bibitem[{{Lis} \& {Schilke}(2003)}]{lisSchilke03}
{Lis}, D.~C. \& {Schilke}, P. 2003, \apjl, 597, L145

\bibitem[{{Madden} {et~al.}(1997){Madden}, {Geis}, {Genzel}, {Nikola},
  {Poglitsch}, {Stacey}, \& {Townes}}]{madden97}
{Madden}, S., {Geis}, N., {Genzel}, R., {et~al.} 1997, in ESA Special
  Publication, Vol. 401, The Far Infrared and Submillimetre Universe., ed.
  A.~{Wilson}, 111

\bibitem[{{Menten} {et~al.}(2007){Menten}, {Reid}, {Forbrich}, \&
  {Brunthaler}}]{menten07}
{Menten}, K.~M., {Reid}, M.~J., {Forbrich}, J., \& {Brunthaler}, A. 2007, \aap,
  474, 515

\bibitem[{{Monje} {et~al.}(2011{\natexlab{a}}){Monje}, {Emprechtinger},
  {Phillips}, {Lis}, {Goldsmith}, {Bergin}, {Bell}, {Neufeld}, \&
  {Sonnentrucker}}]{monje11}
{Monje}, R.~R., {Emprechtinger}, M., {Phillips}, T.~G., {et~al.}
  2011{\natexlab{a}}, \apjl, 734, L23

\bibitem[{{Monje} {et~al.}(2014){Monje}, {Lord}, {Falgarone}, {Lis}, {Neufeld},
  {Phillips}, \& {G{\"u}sten}}]{monje14}
{Monje}, R.~R., {Lord}, S., {Falgarone}, E., {et~al.} 2014, \apj, 785, 22

\bibitem[{{Monje} {et~al.}(2011{\natexlab{b}}){Monje}, {Phillips}, {Peng},
  {Lis}, {Neufeld}, \& {Emprechtinger}}]{monje11_z}
{Monje}, R.~R., {Phillips}, T.~G., {Peng}, R., {et~al.} 2011{\natexlab{b}},
  \apjl, 742, L21

\bibitem[{{Nagy} {et~al.}(2017){Nagy}, {Choi}, {Ossenkopf-Okada}, {van der
  Tak}, {Bergin}, {Gerin}, {Joblin}, {R{\"o}llig}, {Simon}, \&
  {Stutzki}}]{nagy17}
{Nagy}, Z., {Choi}, Y., {Ossenkopf-Okada}, V., {et~al.} 2017, \aap, 599, A22

\bibitem[{{Nagy} {et~al.}(2013){Nagy}, {Van der Tak}, {Ossenkopf}, {Gerin}, {Le
  Petit}, {Le Bourlot}, {Black}, {Goicoechea}, {Joblin}, {R{\"o}llig}, \&
  {Bergin}}]{nagy13}
{Nagy}, Z., {Van der Tak}, F.~F.~S., {Ossenkopf}, V., {et~al.} 2013, \aap, 550,
  A96

\bibitem[{{Neufeld} {et~al.}(2010){Neufeld}, {Sonnentrucker}, {Phillips},
  {Lis}, {de Luca}, {Goicoechea}, {Black}, {Gerin}, {Bell}, {Boulanger},
  {Cernicharo}, {Coutens}, {Dartois}, {Kazmierczak}, {Encrenaz}, {Falgarone},
  {Geballe}, {Giesen}, {Godard}, {Goldsmith}, {Gry}, {Gupta}, {Hennebelle},
  {Herbst}, {Hily-Blant}, {Joblin}, {Ko{\l}os}, {Kre{\l}owski},
  {Mart{\'\i}n-Pintado}, {Menten}, {Monje}, {Mookerjea}, {Pearson}, {Perault},
  {Persson}, {Plume}, {Salez}, {Schlemmer}, {Schmidt}, {Stutzki}, {Teyssier},
  {Vastel}, {Yu}, {Cais}, {Caux}, {Liseau}, {Morris}, \&
  {Planesas}}]{neufeld10}
{Neufeld}, D.~A., {Sonnentrucker}, P., {Phillips}, T.~G., {et~al.} 2010, \aap,
  518, L108

\bibitem[{{Neufeld} \& {Wolfire}(2009)}]{neufeld09}
{Neufeld}, D.~A. \& {Wolfire}, M.~G. 2009, \apj, 706, 1594

\bibitem[{{Neufeld} {et~al.}(1997){Neufeld}, {Zmuidzinas}, {Schilke}, \&
  {Phillips}}]{neufeld97}
{Neufeld}, D.~A., {Zmuidzinas}, J., {Schilke}, P., \& {Phillips}, T.~G. 1997,
  \apjl, 488, L141

\bibitem[{{Ossenkopf} {et~al.}(2013){Ossenkopf}, {R{\"o}llig}, {Neufeld},
  {Pilleri}, {Lis}, {Fuente}, {van der Tak}, \& {Bergin}}]{ossenkopf13}
{Ossenkopf}, V., {R{\"o}llig}, M., {Neufeld}, D.~A., {et~al.} 2013, \aap, 550,
  A57

\bibitem[{{Pabst} {et~al.}(2019){Pabst}, {Higgins}, {Goicoechea}, {Teyssier},
  {Berne}, {Chambers}, {Wolfire}, {Suri}, {Guesten}, {Stutzki}, {Graf},
  {Risacher}, \& {Tielens}}]{pabst19}
{Pabst}, C., {Higgins}, R., {Goicoechea}, J.~R., {et~al.} 2019, \nat, 565, 618

\bibitem[{{Pellegrini} {et~al.}(2009){Pellegrini}, {Baldwin}, {Ferland},
  {Shaw}, \& {Heathcote}}]{Pellegrini2009}
{Pellegrini}, E.~W., {Baldwin}, J.~A., {Ferland}, G.~J., {Shaw}, G., \&
  {Heathcote}, S. 2009, \apj, 693, 285

\bibitem[{{Phillips} {et~al.}(2010){Phillips}, {Bergin}, {Lis}, {Neufeld},
  {Bell}, {Wang}, {Crockett}, {Emprechtinger}, {Blake}, {Caux}, {Ceccarelli},
  {Cernicharo}, {Comito}, {Daniel}, {Dubernet}, {Encrenaz}, {Gerin}, {Giesen},
  {Goicoechea}, {Goldsmith}, {Herbst}, {Joblin}, {Johnstone}, {Langer},
  {Latter}, {Lord}, {Maret}, {Martin}, {Melnick}, {Menten}, {Morris},
  {M{\"u}ller}, {Murphy}, {Ossenkopf}, {Pearson}, {P{\'e}rault}, {Plume},
  {Qin}, {Schilke}, {Schlemmer}, {Stutzki}, {Trappe}, {van der Tak}, {Vastel},
  {Yorke}, {Yu}, {Zmuidzinas}, {Boogert}, {G{\"u}sten}, {Hartogh}, {Honingh},
  {Karpov}, {Kooi}, {Krieg}, \& {Schieder}}]{phillips10}
{Phillips}, T.~G., {Bergin}, E.~A., {Lis}, D.~C., {et~al.} 2010, \aap, 518,
  L109

\bibitem[{{Pilbratt} {et~al.}(2010){Pilbratt}, {Riedinger}, {Passvogel},
  {Crone}, {Doyle}, {Gageur}, {Heras}, {Jewell}, {Metcalfe}, {Ott}, \&
  {Schmidt}}]{pilbratt10}
{Pilbratt}, G.~L., {Riedinger}, J.~R., {Passvogel}, T., {et~al.} 2010, \aap,
  518, L1

\bibitem[{{Qiu} {et~al.}(2018){Qiu}, {Xie}, \& {Zhang}}]{qui18}
{Qiu}, K., {Xie}, Z., \& {Zhang}, Q. 2018, \apj, 855, 48

\bibitem[{{Rangwala} {et~al.}(2011){Rangwala}, {Maloney}, {Glenn}, {Wilson},
  {Rykala}, {Isaak}, {Baes}, {Bendo}, {Boselli}, {Bradford}, {Clements},
  {Cooray}, {Fulton}, {Imhof}, {Kamenetzky}, {Madden}, {Mentuch}, {Sacchi},
  {Sauvage}, {Schirm}, {Smith}, {Spinoglio}, \& {Wolfire}}]{rangwala11}
{Rangwala}, N., {Maloney}, P.~R., {Glenn}, J., {et~al.} 2011, \apj, 743, 94

\bibitem[{{Reese} {et~al.}(2005){Reese}, {Stoecklin}, {Voronin}, \&
  {Rayez}}]{Reese2005}
{Reese}, C., {Stoecklin}, T., {Voronin}, A., \& {Rayez}, J.~C. 2005, \aap, 430,
  1139

\bibitem[{{Roelfsema} {et~al.}(2012){Roelfsema}, {Helmich}, {Teyssier},
  {Ossenkopf}, {Morris}, {Olberg}, {Shipman}, {Risacher}, {Akyilmaz},
  {Assendorp}, {Avruch}, {Beintema}, {Biver}, {Boogert}, {Borys}, {Braine},
  {Caris}, {Caux}, {Cernicharo}, {Coeur-Joly}, {Comito}, {de Lange},
  {Delforge}, {Dieleman}, {Dubbeldam}, {de Graauw}, {Edwards}, {Fich},
  {Flederus}, {Gal}, {di Giorgio}, {Herpin}, {Higgins}, {Hoac}, {Huisman},
  {Jarchow}, {Jellema}, {de Jonge}, {Kester}, {Klein}, {Kooi}, {Kramer},
  {Laauwen}, {Larsson}, {Leinz}, {Lord}, {Lorenzani}, {Luinge}, {Marston},
  {Mart{\'\i}n-Pintado}, {McCoey}, {Melchior}, {Michalska}, {Moreno},
  {M{\"u}ller}, {Nowosielski}, {Okada}, {Orlea{\'n}ski}, {Phillips}, {Pearson},
  {Rabois}, {Ravera}, {Rector}, {Rengel}, {Sagawa}, {Salomons},
  {S{\'a}nchez-Su{\'a}rez}, {Schieder}, {Schl{\"o}der}, {Schm{\"u}lling},
  {Soldati}, {Stutzki}, {Thomas}, {Tielens}, {Vastel}, {Wildeman}, {Xie},
  {Xilouris}, {Wafelbakker}, {Whyborn}, {Zaal}, {Bell}, {Bjerkeli}, {De Beck},
  {Cavali{\'e}}, {Crockett}, {Hily-Blant}, {Kama}, {Kaminski}, {Lefl{\'o}ch},
  {Lombaert}, {de Luca}, {Makai}, {Marseille}, {Nagy}, {Pacheco}, {van der
  Wiel}, {Wang}, \& {Y{\i}ld{\i}z}}]{roelfsema12}
{Roelfsema}, P.~R., {Helmich}, F.~P., {Teyssier}, D., {et~al.} 2012, \aap, 537,
  A17

\bibitem[{{Salgado} {et~al.}(2016){Salgado}, {Bern{\'e}}, {Adams}, {Herter},
  {Keller}, \& {Tielens}}]{salgado16}
{Salgado}, F., {Bern{\'e}}, O., {Adams}, J.~D., {et~al.} 2016, \apj, 830, 118

\bibitem[{{Sch{\"o}ier} {et~al.}(2005){Sch{\"o}ier}, {van der Tak}, {van
  Dishoeck}, \& {Black}}]{schoier05}
{Sch{\"o}ier}, F.~L., {van der Tak}, F.~F.~S., {van Dishoeck}, E.~F., \&
  {Black}, J.~H. 2005, \aap, 432, 369

\bibitem[{{Shaw} {et~al.}(2009){Shaw}, {Ferland}, {Henney}, {Stancil}, {Abel},
  {Pellegrini}, {Baldwin}, \& {van Hoof}}]{Shaw2009}
{Shaw}, G., {Ferland}, G.~J., {Henney}, W.~J., {et~al.} 2009, \apj, 701, 677

\bibitem[{{Shipman} {et~al.}(2017){Shipman}, {Beaulieu}, {Teyssier}, {Morris},
  {Rengel}, {McCoey}, {Edwards}, {Kester}, {Lorenzani}, {Coeur-Joly},
  {Melchior}, {Xie}, {Sanchez}, {Zaal}, {Avruch}, {Borys}, {Braine}, {Comito},
  {Delforge}, {Herpin}, {Hoac}, {Kwon}, {Lord}, {Marston}, {Mueller}, {Olberg},
  {Ossenkopf}, {Puga}, \& {Akyilmaz-Yabaci}}]{russ17}
{Shipman}, R.~F., {Beaulieu}, S.~F., {Teyssier}, D., {et~al.} 2017, \aap, 608,
  A49

\bibitem[{{Simon} {et~al.}(1997){Simon}, {Stutzki}, {Sternberg}, \&
  {Winnewisser}}]{simon97}
{Simon}, R., {Stutzki}, J., {Sternberg}, A., \& {Winnewisser}, G. 1997, \aap,
  327, L9

\bibitem[{{Sim{\'o}n-D{\'\i}az} \& {Stasi{\'n}ska}(2011)}]{simondiaz11}
{Sim{\'o}n-D{\'\i}az}, S. \& {Stasi{\'n}ska}, G. 2011, \aap, 526, A48

\bibitem[{{Sonnentrucker} {et~al.}(2010){Sonnentrucker}, {Neufeld}, {Phillips},
  {Gerin}, {Lis}, {de Luca}, {Goicoechea}, {Black}, {Bell}, {Boulanger},
  {Cernicharo}, {Coutens}, {Dartois}, {Ka{\'z}mierczak}, {Encrenaz},
  {Falgarone}, {Geballe}, {Giesen}, {Godard}, {Goldsmith}, {Gry}, {Gupta},
  {Hennebelle}, {Herbst}, {Hily-Blant}, {Joblin}, {Ko{\l}os}, {Kre{\l}owski},
  {Mart{\'\i}n-Pintado}, {Menten}, {Monje}, {Mookerjea}, {Pearson}, {Perault},
  {Persson}, {Plume}, {Salez}, {Schlemmer}, {Schmidt}, {Stutzki}, {Teyssier},
  {Vastel}, {Yu}, {Caux}, {G{\"u}sten}, {Hatch}, {Klein}, {Mehdi}, {Morris}, \&
  {Ward}}]{sonnentrucker10}
{Sonnentrucker}, P., {Neufeld}, D.~A., {Phillips}, T.~G., {et~al.} 2010, \aap,
  521, L12

\bibitem[{{Stutzki} {et~al.}(1988){Stutzki}, {Stacey}, {Genzel}, {Graf},
  {Harris}, {Jaffe}, {Lugten}, \& {Poglitsch}}]{stutzki88}
{Stutzki}, J., {Stacey}, G.~J., {Genzel}, R., {et~al.} 1988, {Physical
  conditions in molecular clouds derived from sub-mm and far-IR spectroscopic
  observations}, Tech. rep.

\bibitem[{{Tauber} {et~al.}(1994){Tauber}, {Tielens}, {Meixner}, \&
  {Goldsmith}}]{tauber94}
{Tauber}, J.~A., {Tielens}, A.~G.~G.~M., {Meixner}, M., \& {Goldsmith}, P.~F.
  1994, \apj, 422, 136

\bibitem[{Thummel {et~al.}(1992)Thummel, Nesbet, \& Peyerimhoff}]{thummel92}
Thummel, H.~T., Nesbet, R.~K., \& Peyerimhoff, S.~D. 1992, Journal of Physics
  B: Atomic, Molecular and Optical Physics, 25, 4553

\bibitem[{{Tielens} \& {Hollenbach}(1985)}]{tielens85}
{Tielens}, A.~G.~G.~M. \& {Hollenbach}, D. 1985, \apj, 291, 747

\bibitem[{{Tielens} {et~al.}(1993){Tielens}, {Meixner}, {van der Werf},
  {Bregman}, {Tauber}, {Stutzki}, \& {Rank}}]{tielens93}
{Tielens}, A.~G.~G.~M., {Meixner}, M.~M., {van der Werf}, P.~P., {et~al.} 1993,
  Science, 262, 86

\bibitem[{{van der Tak}(2012b)}]{floris12b}
{van der Tak}, F.~F.~S. 2012b, Philosophical Transactions of the Royal Society
  of London Series A, 370, 5186

\bibitem[{{van der Tak} {et~al.}(2007){van der Tak}, {Black}, {Sch{\"o}ier},
  {Jansen}, \& {van Dishoeck}}]{floris07}
{van der Tak}, F.~F.~S., {Black}, J.~H., {Sch{\"o}ier}, F.~L., {Jansen}, D.~J.,
  \& {van Dishoeck}, E.~F. 2007, \aap, 468, 627

\bibitem[{{van der Tak} {et~al.}(2012a){van der Tak}, {Ossenkopf}, {Nagy},
  {Faure}, {R{\"o}llig}, \& {Bergin}}]{floris12a}
{van der Tak}, F.~F.~S., {Ossenkopf}, V., {Nagy}, Z., {et~al.} 2012a, \aap,
  537, L10

\bibitem[{{van der Werf} {et~al.}(2013){van der Werf}, {Goss}, \&
  {O'Dell}}]{vanderwerf2013}
{van der Werf}, P.~P., {Goss}, W.~M., \& {O'Dell}, C.~R. 2013, \apj, 762, 101

\bibitem[{{van der Werf} {et~al.}(2010){van der Werf}, {Isaak}, {Meijerink},
  {Spaans}, {Rykala}, {Fulton}, {Loenen}, {Walter}, {Wei{\ss}}, {Armus},
  {Fischer}, {Israel}, {Harris}, {Veilleux}, {Henkel}, {Savini}, {Lord},
  {Smith}, {Gonz{\'a}lez-Alfonso}, {Naylor}, {Aalto}, {Charmand aris},
  {Dasyra}, {Evans}, {Gao}, {Greve}, {G{\"u}sten}, {Kramer},
  {Mart{\'\i}n-Pintado}, {Mazzarella}, {Papadopoulos}, {Sanders}, {Spinoglio},
  {Stacey}, {Vlahakis}, {Wiedner}, \& {Xilouris}}]{vanderwerf10}
{van der Werf}, P.~P., {Isaak}, K.~G., {Meijerink}, R., {et~al.} 2010, \aap,
  518, L42

\bibitem[{{van der Wiel} {et~al.}(2016){van der Wiel}, {Naylor}, {Makiwa},
  {Satta}, \& {Abergel}}]{vanderwiel16}
{van der Wiel}, M.~H.~D., {Naylor}, D.~A., {Makiwa}, G., {Satta}, M., \&
  {Abergel}, A. 2016, \aap, 593, A37

\bibitem[{{Walmsley} {et~al.}(2000){Walmsley}, {Natta}, {Oliva}, \&
  {Testi}}]{walmsley00}
{Walmsley}, C.~M., {Natta}, A., {Oliva}, E., \& {Testi}, L. 2000, \aap, 364,
  301

\bibitem[{{Wang} {et~al.}(1993){Wang}, {Jaffe}, {Evans}, {Hayashi},
  {Tatematsu}, \& {Zhou}}]{wang93}
{Wang}, Y., {Jaffe}, D.~T., {Evans}, Neal~J., I., {et~al.} 1993, \apj, 419, 707

\bibitem[{{Weilbacher} {et~al.}(2015){Weilbacher}, {Monreal-Ibero},
  {Kollatschny}, {Ginsburg}, {McLeod}, {Kamann}, {Sandin}, {Palsa}, {Wisotzki},
  {Bacon}, {Selman}, {Brinchmann}, {Caruana}, {Kelz}, {Martinsson},
  {P{\'e}contal-Rousset}, {Richard}, \& {Wendt}}]{orionbarOI}
{Weilbacher}, P.~M., {Monreal-Ibero}, A., {Kollatschny}, W., {et~al.} 2015,
  \aap, 582, A114

\bibitem[{{Wolfire} {et~al.}(2003){Wolfire}, {McKee}, {Hollenbach}, \&
  {Tielens}}]{wolfire03}
{Wolfire}, M.~G., {McKee}, C.~F., {Hollenbach}, D., \& {Tielens}, A.~G.~G.~M.
  2003, \apj, 587, 278

\bibitem[{{Yang} {et~al.}(2015){Yang}, {Walker}, {Forrey}, {Stancil}, \&
  {Balakrishnan}}]{yang15}
{Yang}, B., {Walker}, K.~M., {Forrey}, R.~C., {Stancil}, P.~C., \&
  {Balakrishnan}, N. 2015, \aap, 578, A65

\bibitem[{{Young Owl} {et~al.}(2000){Young Owl}, {Meixner}, {Wolfire},
  {Tielens}, \& {Tauber}}]{youngowl00}
{Young Owl}, R.~C., {Meixner}, M.~M., {Wolfire}, M., {Tielens}, A.~G.~G.~M., \&
  {Tauber}, J. 2000, \apj, 540, 886

\end{thebibliography}
\begin{appendix}

\section{SEDs of Three Positions in the HF map}\label{app1}
To determine the spatial distribution of dust temperature and column density in the Orion Bar, we use \textit{Herschel} PACS ($70\ \mu$m and $160\ \mu$m) and SPIRE ($250\ \mu$m, $350\ \mu$m, and $500\ \mu$m) maps. All maps are convolved to the SPIRE $500\ \mu$m beam size of $39^{\arcsec}$ FWHM. To construct the SED of the Orion Bar, we choose 3 positions within the HF integrated intensity map (see Figure~\ref{fig:integratedmap}). The flux densities are modeled as a modified blackbody,
\begin{equation*}
I(\lambda) = B(\lambda, T_d)~\tau_0~\Bigg(\frac{\lambda_0}{\lambda}\Bigg)^\beta
\end{equation*}
Here, $T_\mathrm{d}$ denotes the effective dust temperature, $\tau_0$ the dust optical depth at the reference wavelength $\lambda_0$, and $\beta$ the dust grain opacity index. The reference wavelength ($\lambda_0$) is the position of the HF $1232.476\ \mathrm{GHz}$. $T_{d}$ and $\tau_0$ are free parameters. Here, we assume that the dust emission is optically thin. The dust emissivity index ($\beta$) is fixed at 1.7 in all models \citep{Arab2012}. We fit the fluxes with a modified blackbody at three different positions. In front of the Bar, position 1, the fitted temperature is $49\ \mathrm{K}$ and it decreases slightly to $43\ \mathrm{K}$ in the Orion Bar, position 2. The temperature in the deeper cloud, position 3, is similar to the temperature in the Bar. 

We run two RADEX models at the HF peak, position 2. In the first model, we run RADEX considering only CMB emission. For a gas kinetic temperature of $120\ \mathrm{K}$, this model predicts an intensity for the HF \textit{J} = 1$\,\to\,$0 line of $1.97\ \mathrm{K}$. The second model where we only added the IR radiation field coming from dust at $50\ \mathrm{K}$ to CMB also predicts same intensity for the HF \textit{J} = 1$\,\to\,$0 line, that is, $1.97\ \mathrm{K}$. The RADEX models show that FIR pumping by $50\ \mathrm{K}$ warm dust is not important. More detailed models have been developed by \citet{Shaw2009} involving detailed temperature profile, but we feel that this is outside the scope of this paper. We elected a more straightforward approach by \citet{salgado16}. Following \citet{salgado16}, dust IR emission optically thin at all positions. Subsequently, CMB emission is only used in the models.

\begin{figure}[ht]
    \centering
    \includegraphics[width=1.0\hsize]{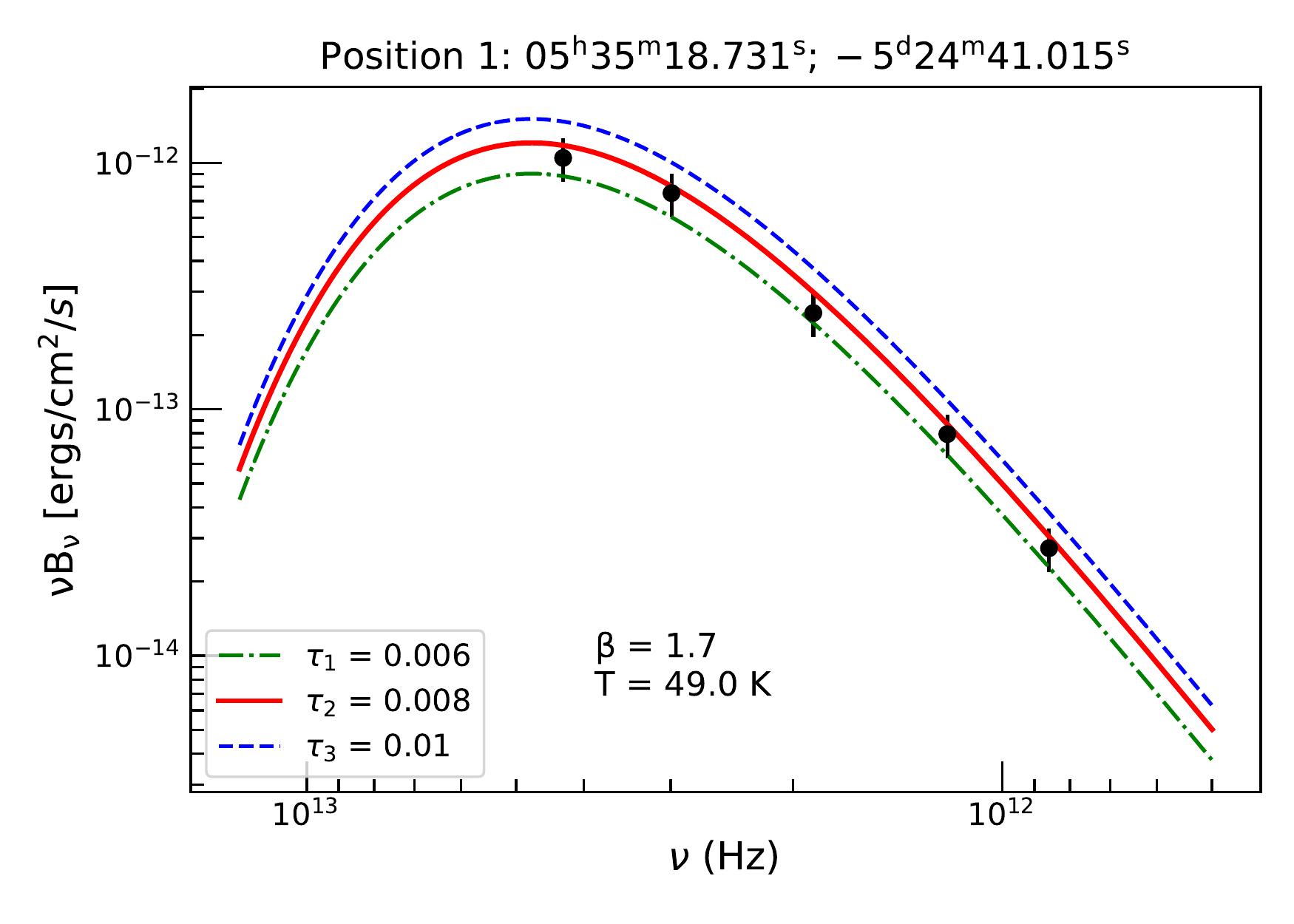}
    \includegraphics[width=1.0\hsize]{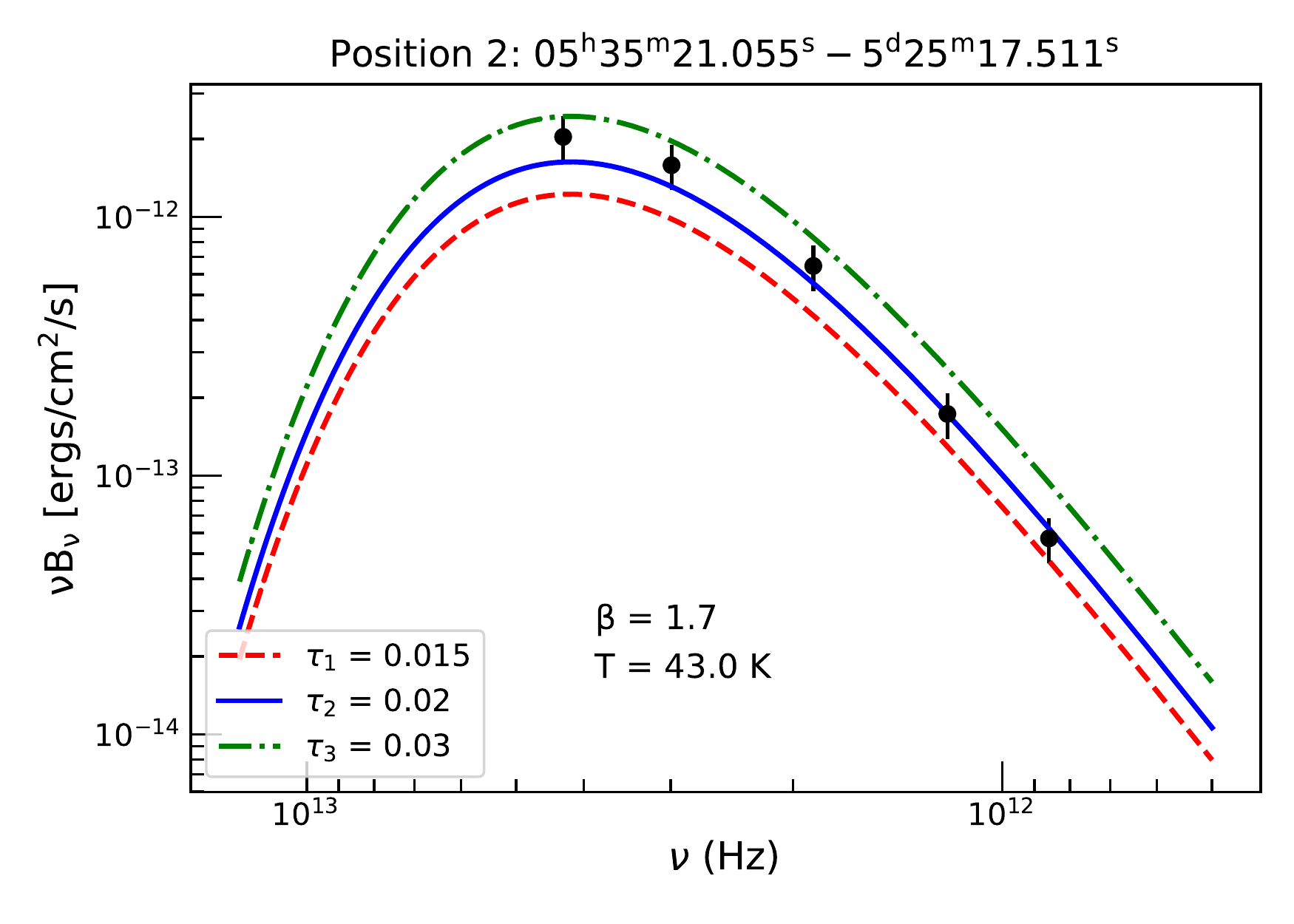}
    \includegraphics[width=1.0\hsize]{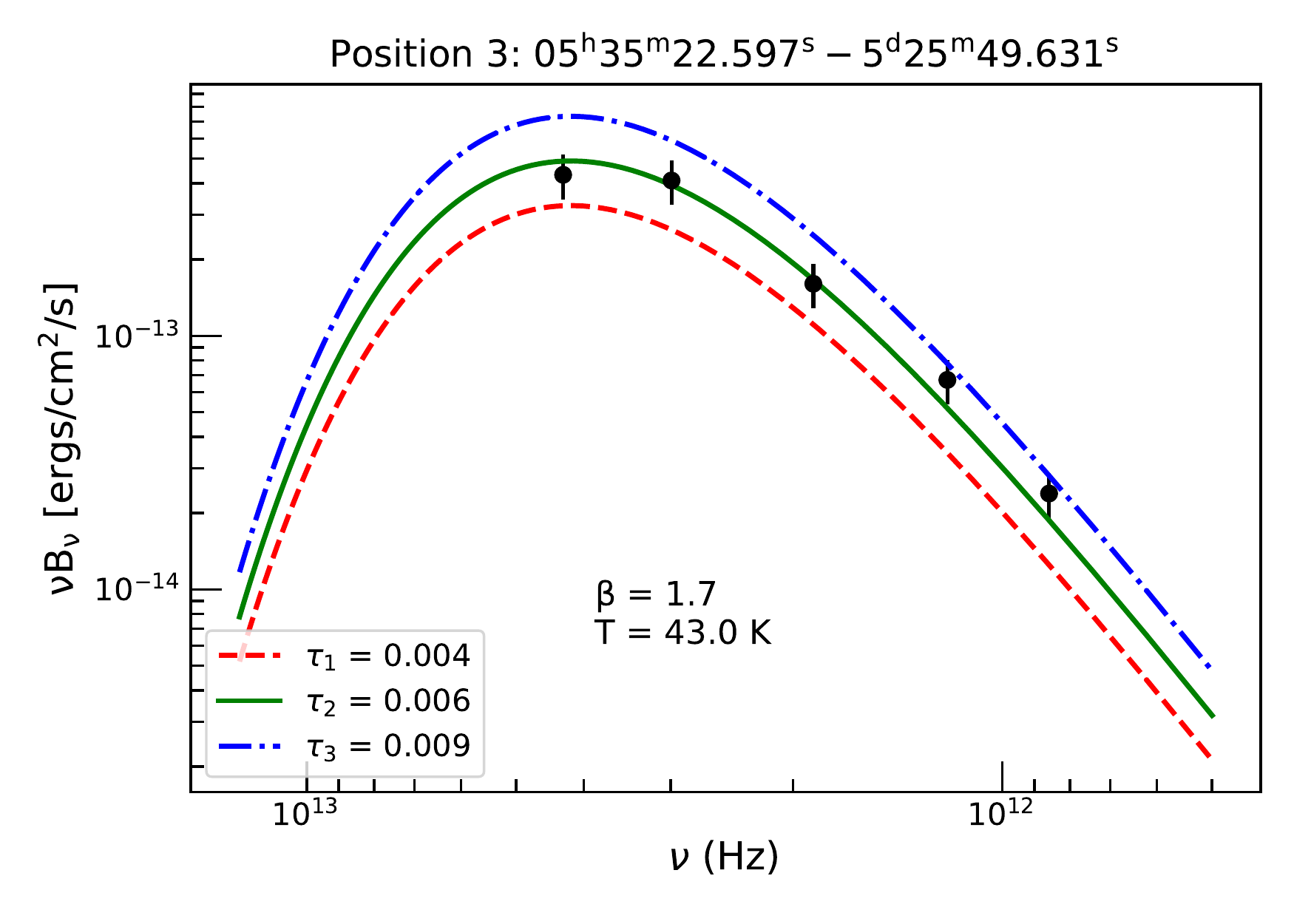}
\caption{SED of three positions within the HF map as labeled in the Figure~\ref{fig:integratedmap}.}
\label{fig:HFSEDs}
\end{figure}

\newpage

\begin{table*}
\caption{Line parameters and column densities of the spectrum at found each pixel.} 
\label{table:lineparameters}
\centering                  
\begin{tabular}{c c c c c c c c c}       
\hline\hline             
R.A.(J2000) & Dec.(J2000) & $\int T_\mathrm{mb} \Delta$V  & $V_\mathrm{LSR}$          & $\Delta$V        & $T_\mathrm{mb}$ &  $N_\mathrm{col}$ & $N_\mathrm{col}$ (50~K) & N$_{col}$ (35 K)   \\
(h:m:s) & ($^{\degr}$:$^{\arcmin}$:$^{\arcsec}$)   &  [K km s$^{-1}$]   & [km~s$^{-1}$]      & [km s$^{-1}$]    & [K]      &  10$^{15}$~[cm$^{-2}$] & 10$^{15}$~[cm$^{-2}$] & 10$^{15}$~[cm$^{-2}$] \\
\hline
 5:35:19.6    & -5:24:24.6     & 3.62 $\pm$ 0.24    &  8.97  $\pm$ 0.09  & 2.63 $\pm$ 0.19  & 1.29        & 0.50 & 1.10 & 1.11 \\
 5:35:19.1    & -5:24:24.6     & 3.62 $\pm$ 0.24    &  8.95  $\pm$ 0.10  & 2.95 $\pm$ 0.21  & 1.15        & 0.48 & 1.09 & 1.10 \\
 5:35:18.5    & -5:24:24.6     & 4.04 $\pm$ 0.27    &  8.27  $\pm$ 0.13  & 4.22 $\pm$ 0.33  & 0.90        & 0.53 & 1.20 & 1.21 \\
 5:35:19.6    & -5:24:32.8     & 3.44 $\pm$ 0.18    &  8.88  $\pm$ 0.08  & 2.89 $\pm$ 0.17  & 0.12        & 0.46 & 1.04 & 1.04 \\
 5:35:19.1    & -5:24:32.8     & 3.55 $\pm$ 0.19    &  8.88  $\pm$ 0.08  & 3.16 $\pm$ 0.18  & 1.05        & 0.47 & 1.07 & 1.07 \\
 5:35:18.5    & -5:24:32.8     & 4.04 $\pm$ 0.26    &  8.35  $\pm$ 0.14  & 4.26 $\pm$ 0.30  & 0.89        & 0.56 & 1.20 & 1.21 \\ 
 5:35:18.0    & -5:24:32.8     & 4.02 $\pm$ 0.28    &  8.31  $\pm$ 0.16  & 4.32 $\pm$ 0.34  & 0.87        & 0.52 & 1.20 & 1.21 \\ 
 5:35:20.2    & -5:24:41.5     & 3.95 $\pm$ 0.22    &  9.41  $\pm$ 0.11  & 4.05 $\pm$ 0.25  & 0.91        & 0.52 & 1.18 & 1.19 \\
 5:35:19.6    & -5:24:41.5     & 3.59 $\pm$ 0.16    &  8.87  $\pm$ 0.08  & 3.44 $\pm$ 0.16  & 0.98        & 0.47 & 1.07 & 1.08 \\
 5:35:19.1    & -5:24:41.5     & 3.73 $\pm$ 0.13    &  8.54  $\pm$ 0.06  & 3.57 $\pm$ 0.13  & 0.98        & 0.49 & 1.11 & 1.12 \\ 
 5:35:18.5    & -5:24:41.5     & 3.94 $\pm$ 0.15    &  8.37  $\pm$ 0.07  & 3.79 $\pm$ 0.15  & 0.97        & 0.52 & 1.18 & 1.19 \\ 
 5:35:18.0    & -5:24:41.5     & 4.09 $\pm$ 0.20    &  8.33  $\pm$ 0.11  & 4.16 $\pm$ 0.22  & 0.92        & 0.54 & 1.22 & 1.23 \\ 
 5:35:20.8    & -5:24:50.4     & 2.76 $\pm$ 0.17    &  9.87  $\pm$ 0.07  & 2.60 $\pm$ 0.19  & 1.00        & 0.36 & 0.83 & 0.83 \\
 5:35:20.2    & -5:24:50.4     & 3.15 $\pm$ 0.17    &  9.54  $\pm$ 0.07  & 2.89 $\pm$ 0.21  & 1.02        & 0.41 & 0.94 & 0.95 \\
 5:35:19.6    & -5:24:50.4     & 3.73 $\pm$ 0.17    &  9.33  $\pm$ 0.08  & 3.73 $\pm$ 0.19  & 0.93        & 0.49 & 1.11 & 1.12 \\
 5:35:19.1    & -5:24:50.4     & 4.14 $\pm$ 0.15    &  8.87  $\pm$ 0.08  & 4.35 $\pm$ 0.18  & 0.89        & 0.54 & 1.23 & 1.24 \\
 5:35:18.5    & -5:24:50.4     & 4.62 $\pm$ 0.15    &  8.26  $\pm$ 0.07  & 4.14 $\pm$ 0.15  & 1.05        & 0.61 & 1.38 & 1.40 \\
 5:35:18.0    & -5:24:50.4     & 4.98 $\pm$ 1.22    &  7.71  $\pm$ 0.62  & 3.68 $\pm$ 1.30  & 1.27        & 0.66 & 1.51 & 1.52 \\
 5:35:21.4    & -5:24:59.0     & 7.12 $\pm$ 0.35    &  9.67  $\pm$ 0.11  & 4.73 $\pm$ 0.28  & 1.41        & 0.97 & 2.17 & 2.19 \\
 5:35:20.8    & -5:24:59.0     & 4.55 $\pm$ 0.17    &  9.87  $\pm$ 0.07  & 3.62 $\pm$ 0.15  & 1.18        & 0.61 & 1.37 & 1.39 \\
 5:35:20.2    & -5:24:59.0     & 3.17 $\pm$ 0.14    &  9.85  $\pm$ 0.07  & 3.40 $\pm$ 0.17  & 0.87        & 0.41 & 0.94 & 0.95 \\ 
 5:35:19.6    & -5:24:59.0     & 3.52 $\pm$ 0.11    &  9.50  $\pm$ 0.06  & 3.92 $\pm$ 0.13  & 0.84        & 0.46 & 1.05 & 1.05 \\ 
 5:35:19.1    & -5:24:59.0     & 3.74 $\pm$ 0.20    &  9.10  $\pm$ 0.10  & 3.87 $\pm$ 0.25  & 0.90        & 0.49 & 1.11 & 1.12 \\ 
 5:35:18.5    & -5:24:59.0     & 4.42 $\pm$ 0.23    &  9.02  $\pm$ 0.12  & 4.69 $\pm$ 0.29  & 0.88        & 0.58 & 1.31 & 1.33 \\ 
 5:35:21.4    & -5:25:07.3     & 8.61 $\pm$ 0.17    &  10.50 $\pm$ 0.04  & 3.86 $\pm$ 0.08  & 2.09        & 1.23 & 2.73 & 2.75 \\
 5:35:20.8    & -5:25:07.3     & 7.09 $\pm$ 0.16    &  10.27 $\pm$ 0.05  & 4.17 $\pm$ 0.10  & 1.60        & 0.97 & 2.19 & 2.21 \\
 5:35:20.8    & -5:25:07.3     & 5.65 $\pm$ 0.14    &  9.98  $\pm$ 0.05  & 4.28 $\pm$ 0.12  & 1.24        & 0.76 & 1.71 & 1.73 \\
 5:35:19.6    & -5:25:07.3     & 4.08 $\pm$ 0.19    &  9.86  $\pm$ 0.09  & 4.05 $\pm$ 0.20  & 0.94        & 0.53 & 1.22 & 1.23 \\
 5:35:19.1    & -5:25:07.3     & 3.76 $\pm$ 0.13    &  9.34  $\pm$ 0.08  & 4.27 $\pm$ 0.16  & 0.82        & 0.49 & 1.11 & 1.12 \\
 5:35:22.0    & -5:25:16.0     & 8.66 $\pm$ 0.20    &  10.57 $\pm$ 0.05  & 3.88 $\pm$ 0.09  & 2.09        & 1.24 & 2.74 & 2.76 \\
 5:35:21.4    & -5:25:16.0     & 9.35 $\pm$ 0.19    &  10.53 $\pm$ 0.04  & 3.87 $\pm$ 0.07  & 2.27        & 1.35 & 2.99 & 3.01 \\
 5:35:20.8    & -5:25:16.0     & 8.93 $\pm$ 0.14    &  10.52 $\pm$ 0.03  & 3.87 $\pm$ 0.07  & 2.17        & 1.28 & 2.84 & 2.86 \\
 5:35:20.2    & -5:25:16.0     & 8.51 $\pm$ 0.20    &  10.20 $\pm$ 0.05  & 4.41 $\pm$ 0.11  & 1.80        & 1.19 & 2.66 & 2.68 \\
 5:35:19.6    & -5:25:16.0     & 6.51 $\pm$ 0.22    &  10.01 $\pm$ 0.08  & 4.77 $\pm$ 0.18  & 1.28        & 0.87 & 1.98 & 1.99 \\ 
 5:35:22.5    & -5:25:25.0     & 5.34 $\pm$ 0.20    &  10.50 $\pm$ 0.08  & 4.00 $\pm$ 0.15  & 1.25        & 0.71 & 1.62 & 1.63 \\
 5:35:22.0    & -5:25:25.0     & 6.06 $\pm$ 0.18    &  10.62 $\pm$ 0.05  & 3.73 $\pm$ 0.11  & 1.52        & 0.83 & 1.87 & 1.88 \\
 5:35:21.4    & -5:25:25.0     & 7.79 $\pm$ 0.15    &  10.58 $\pm$ 0.04  & 3.88 $\pm$ 0.08  & 1.89        & 1.09 & 2.44 & 2.46 \\ 
 5:35:20.8    & -5:25:25.0     & 8.72 $\pm$ 0.16    &  10.57 $\pm$ 0.04  & 3.94 $\pm$ 0.08  & 2.08        & 1.24 & 2.76 & 2.78 \\ 
 5:35:20.2    & -5:25:25.0     & 9.17 $\pm$ 0.21    &  10.52 $\pm$ 0.05  & 4.19 $\pm$ 0.10  & 2.05        & 1.30 & 2.90 & 2.92 \\
 5:35:19.6    & -5:25:25.0     & 9.14 $\pm$ 0.43    &  10.20 $\pm$ 0.11  & 4.27 $\pm$ 0.21  & 2.01        & 1.29 & 2.88 & 2.90 \\
 5:35:22.5    & -5:25:33.7     & 4.43 $\pm$ 0.20    &  10.40 $\pm$ 0.09  & 4.13 $\pm$ 0.22  & 1.00        & 0.58 & 1.33 & 1.34 \\
 5:35:22.0    & -5:25:33.7     & 4.99 $\pm$ 0.16    &  10.51 $\pm$ 0.06  & 3.99 $\pm$ 0.13  & 1.17        & 0.66 & 1.51 & 1.52 \\
 5:35:21.4    & -5:25:33.7     & 5.79 $\pm$ 0.12    &  10.48 $\pm$ 0.04  & 3.94 $\pm$ 0.09  & 1.38        & 0.78 & 1.77 & 1.78 \\ 
 5:35:20.8    & -5:25:33.7     & 7.14 $\pm$ 0.17    &  10.49 $\pm$ 0.05  & 3.95 $\pm$ 0.10  & 1.70        & 0.99 & 2.22 & 2.23 \\ 
 5:35:20.2    & -5:25:33.7     & 8.17 $\pm$ 0.21    &  10.55 $\pm$ 0.05  & 3.95 $\pm$ 0.11  & 1.94        & 1.15 & 2.57 & 2.58 \\
 5:35:23.1    & -5:25:42.0     & 2.85 $\pm$ 0.20    &  10.15 $\pm$ 0.12  & 3.39 $\pm$ 0.26  & 0.79        & 0.37 & 0.84 & 1.85 \\
 5:35:22.5    & -5:25:42.0     & 3.36 $\pm$ 0.17    &  10.29 $\pm$ 0.10  & 3.77 $\pm$ 0.21  & 0.84        & 0.44 & 0.99 & 1.00 \\
 5:35:22.0    & -5:25:42.0     & 3.35 $\pm$ 0.13    &  10.10 $\pm$ 0.08  & 4.01 $\pm$ 0.17  & 0.78        & 0.43 & 0.99 & 1.00 \\ 
 5:35:21.4    & -5:25:42.0     & 4.17 $\pm$ 0.16    &  10.39 $\pm$ 0.09  & 4.39 $\pm$ 0.18  & 0.89        & 0.54 & 1.24 & 1.25 \\ 
 5:35:20.8    & -5:25:42.0     & 5.40 $\pm$ 0.16    &  10.30 $\pm$ 0.06  & 4.26 $\pm$ 0.14  & 1.19        & 0.72 & 1.63 & 1.64 \\
 5:35:23.7    & -5:25:50.7     & 2.17 $\pm$ 0.33    &  9.63  $\pm$ 0.20  & 2.71 $\pm$ 0.57  & 0.75        & 0.28 & 0.64 & 0.64 \\
 5:35:23.1    & -5:25:50.7     & 2.75 $\pm$ 0.22    &  9.73  $\pm$ 0.14  & 3.87 $\pm$ 0.38  & 0.68        & 0.35 & 0.81 & 0.81 \\
 5:35:22.5    & -5:25:50.7     & 2.24 $\pm$ 0.13    &  9.83  $\pm$ 0.10  & 3.51 $\pm$ 0.22  & 0.60        & 0.29 & 0.66 & 0.66 \\
 5:35:22.0    & -5:25:50.7     & 2.37 $\pm$ 0.16    &  10.07 $\pm$ 0.13  & 3.77 $\pm$ 0.29  & 0.59        & 0.30 & 0.69 & 0.70 \\ 
 5:35:21.4    & -5:25:50.7     & 2.35 $\pm$ 0.17    &  9.91  $\pm$ 0.15  & 3.92 $\pm$ 0.29  & 0.56        & 0.30 & 0.69 & 0.69 \\ 
 5:35:23.7    & -5:25:59.7     & 1.82 $\pm$ 0.35    &  9.52  $\pm$ 0.24  & 2.24 $\pm$ 0.64  & 0.76        & 0.23 & 0.54 & 0.54 \\
 
\hline
\end{tabular}
\end{table*}

\newpage

\begin{table*}
\caption{continued} 
\centering                  
\begin{tabular}{c c c c c c c c c}       
\hline\hline             
R.A.(J2000) & Dec.(J2000) &$\int T_\mathrm{mb} \Delta$V  & $V_\mathrm{LSR}$          & $\Delta$V        & $T_\mathrm{mb}$ &  $N_\mathrm{col}$ & $N_\mathrm{col}$ (50~K) & $N_{col}$ (35 K)   \\
(h:m:s) & ($^{\degr}$:$^{\arcmin}$:$^{\arcsec}$)   &  [K km s$^{-1}$]   & [km~s$^{-1}$]      & [km s$^{-1}$]    & [K]      &  10$^{15}$~[cm$^{-2}$] & 10$^{15}$~[cm$^{-2}$] & 10$^{15}$~[cm$^{-2}$] \\
\hline
 5:35:23.1    & -5:25:59.7     & 1.75 $\pm$ 0.31    &  9.52  $\pm$ 0.21  & 2.26 $\pm$ 0.65  & 0.73        & 0.22 & 0.52 & 0.52 \\
 5:35:22.5    & -5:25:59.7     & 1.91 $\pm$ 0.17    &  9.35  $\pm$ 0.15  & 3.40 $\pm$ 0.35  & 0.53        & 0.24 & 0.56 & 0.56 \\ 
 5:35:22.0    & -5:25:59.7     & 1.35 $\pm$ 0.15    &  9.71  $\pm$ 0.17  & 2.87 $\pm$ 0.38  & 0.44        & 0.17 & 0.39 & 0.39 \\ 
 5:35:21.4    & -5:25:59.7     & 2.33 $\pm$ 0.28    &  9.82  $\pm$ 0.20  & 3.82 $\pm$ 0.42  & 0.57        & 0.30 & 0.68 & 0.69 \\ 
 5:35:23.1    & -5:26:07.8     & 2.06 $\pm$ 0.27    &  9.97  $\pm$ 0.26  & 3.48 $\pm$ 0.51  & 0.56        & 0.26 & 0.60 & 0.60 \\
 5:35:22.5    & -5:26:07.8     & 2.08 $\pm$ 0.28    &  9.35  $\pm$ 0.20  & 3.40 $\pm$ 0.49  & 0.58        & 0.27 & 0.61 & 0.61 \\
 5:35:22.0    & -5:26:07.8     & 1.78 $\pm$ 0.25    &  9.18  $\pm$ 0.26  & 3.55 $\pm$ 0.52  & 0.47        & 0.22 & 0.52 & 0.52 \\

\hline
\end{tabular}
\end{table*}

\end{appendix}

\end{document}